\documentclass[aps,prb,groupedaddress,amssym,floatfix,notitlepage,eqsecnum]{revtex4-1}

\usepackage{amsfonts}
\usepackage{amsmath}
\usepackage{amssymb}
\usepackage{graphicx}
\begin{document}


\title{A fermionic approach to tunneling through junctions of multiple quantum wires}
\author{Zheng Shi}
\author{Ian Affleck}
\affiliation{Department of Physics and Astronomy, University of British Columbia,
Vancouver, BC, Canada V6T 1Z1}





\date{\today}

\begin{abstract}
Junctions of multiple one-dimensional quantum wires of interacting electrons have received considerable theoretical attention as a basic constituent of quantum circuits.  While results have been obtained on these models using bosonization and Density Matrix Renormalization Group (DMRG) methods, another powerful technique is based on direct perturbation theory in the bulk interactions, combined with the Renormalization Group (RG) and summed in the Random Phase Approximation (RPA). This technique has so far only been applied to the case where finite length interacting wires are attached to non-interacting Fermi liquid leads. We reformulate it in terms of the single-particle S-matrix, formally unifying treatments of junctions of different numbers of leads, and extend this method to cover the case of infinite length interacting leads obtaining results on 2-lead and 3-lead junctions in good agreement with previous bosonization and DMRG results.
\end{abstract}



\maketitle

\section{Introduction and conclusion\label{sec:intro}}

A class of powerful theoretical approaches to junctions models the quantum
wires as conformally invariant bulk Tomonaga-Luttinger liquids
(TLL).\cite{PhysRevLett.68.1220,*PhysRevB.46.15233,PhysRevB.47.4631,*PhysRevB.47.3827,Wong1994403,1742-5468-2006-02-P02008,PhysRevB.77.155422,PhysRevB.86.075451} In the spirit of boundary conformal field theory, at low energies the
junction with its boundary operators should eventually renormalize to
conformally invariant boundary conditions. Possible fixed points of the
renormalization group (RG) flow are then postulated, and their various
properties, such as zero-temperature conductance and operator scaling dimensions, are explored. Details of the RG flow, however, are largely open to conjecture except in the vicinity of these fixed points. These
approaches are often consolidated with the technique of bosonization, as the
elementary excitations of TLLs are bosonic in nature, and various boundary
conditions imposed by the junction are often conveniently expressed in bosonic
field variables.

Alternate formalisms have been independently developed in the language of fermions. Consider first the junction system without bulk electron-electron interaction in the quantum wires; we further ignore local interactions so that this system becomes completely non-interacting. (For a free-fermion system, unless localized discrete states exist as is true for resonant tunneling, local interactions near the junction are irrelevant in the RG sense and do not affect leading order low-energy physics. We do not treat the case with discrete states in this paper.) A single-particle S-matrix determines the scattering basis, a new set of single-particle states which diagonalizes the non-interacting system. The bulk electron-electron interaction is then reintroduced and handled by perturbation theory, whose infrared divergence is resummed in an RG procedure. The S-matrix elements are now scale-dependent coupling constants. In the simplest approximation scheme, both the renormalized single-particle self-energy and the renormalized two-particle vertex depend only on the renormalized S-matrix at every point in the RG flow; the S-matrix elements (or the transmission probabilities) are thus treated as the only running coupling constants of the theory, and by solving their RG equations we gain information about the conductance. This scheme was first adopted by Refs.~\onlinecite{PhysRevLett.71.3351,*PhysRevB.49.1966,PhysRevB.66.165327,*PhysRevB.70.085318} to the first order in interaction; various generalizations include resonant tunneling with an energy-dependent S-matrix,\cite{PhysRevLett.91.126804,PhysRevB.68.035421} second order perturbation theory in interaction,\cite{PhysRevLett.105.266404,PhysRevB.86.035137} random phase approximation (RPA) in interaction,\cite{0295-5075-82-2-27001,PhysRevB.80.045109,LithJPhys.52.2353,PhysRevB.84.155426,PhysRevB.88.075131} superconducting junctions,\cite{PhysRevLett.97.237006,PhysRevB.77.155418,*PhysRevB.78.205421,*PhysRevB.79.155416} non-equilibrium transport,\cite{PhysRevB.90.245414} and refermionization of fixed points proposed by the bosonic theory,\cite{PhysRevB.92.125138} to list a few. In particular, it has been found that the RPA in the Tomonaga-Luttinger model with a linear dispersion reproduces various scaling dimensions of the conductance known from bosonic methods. (The term RPA has been used interchangeably with ``ladder approximation'' in Refs.~\onlinecite{0295-5075-82-2-27001,PhysRevB.80.045109,LithJPhys.52.2353,PhysRevB.84.155426,PhysRevB.88.075131}.) An improved approximation, known as the functional RG method,\cite{PhysRevB.65.045318,*JLowTempPhys.126.1147,*PhysRevB.67.193303,EurophysLett.64.769,*PhysRevB.71.155401,PhysRevLett.94.136405,*PhysRevB.71.205327} explicitly studies the flows of single-particle self-energy and the two-particle vertex. Despite its basis on perturbation theory in interaction, the functional RG shows excellent agreement with analytic results for two-lead junctions at Luttinger parameter $K=1/2$, and with numerical density matrix renormalization group (DMRG) data up to fairly strong interactions. The merit of these formalisms are that the crossover behavior between different fixed points can, in principle, be found to any order in interaction. Nevertheless, when the interaction becomes sufficiently strong in a junction of three wires (a ``Y-junction''), the fixed points and the RG flow predicted by the RPA fermionic approach and the bosonic approach begin to differ qualitatively. Also a careful analysis reveals that, in the RPA approach, the $\beta$ function of the S-matrix beyond one-loop order contains non-universal terms\cite{0295-5075-82-2-27001,PhysRevB.80.045109,PhysRevB.84.155426} which depend on the precise cutoff scheme of the theory, and may potentially change its predictions.

To our knowledge, many aspects of the junction problem have not been explored in the fermionic formalism. One such example is the well-known distinction between a semi-infinite TLL wire and a finite TLL wire connected to a Fermi liquid (FL) reservoir.\cite{PhysRevB.26.7063,PhysRevLett.68.1220,PhysRevLett.73.468,PhysRevB.52.R5539,PhysRevB.52.R8666,PhysRevB.52.R17040,*PhysRevB.55.R7331,PhysRevB.54.R5239,JPSJ.65.30,PhysRevB.54.R14265,PhysRevLett.87.026801,*PhysRevB.65.195304,PhysRevB.66.035313,*PhysRevB.68.205110,PhysRevB.74.085301,PhysRevB.83.115330} There have been controversies on the nature of the conductance measured in a realistic experimental setup.\cite{PhysRevB.66.035313,PhysRevB.83.115330} However, if we consider the linear DC response to an externally applied bias voltage (rather than the total potential drop which includes induced polarization of the electronic density),\cite{JPSJ.65.30,PhysRevB.54.R14265} then it has been predicted that the corresponding linear response coefficients (which we henceforth refer to as the ``conductances'') with and without the FL reservoir (or ``lead'') are generally different. (These coefficients are well-defined and can be studied numerically.)\cite{PhysRevB.85.045120} For instance, the conductance of a finite TLL wire attached to FL leads on both sides is $e^{2}/h$, irrespective of the interaction strength; on the other hand, the conductance of an infinite spinless TLL wire is $Ke^{2}/h$, where $K$ is the Luttinger parameter. The Landauer formula based on a perfectly transmitting S-matrix alone cannot recover the $Ke^{2}/h$ result. In existing literature employing the fermionic formalism, the case of FL leads has been well studied, but the effects of TLL leads on the conductance are not discussed.

The reasons are twofold for our interest in the effects of TLL leads on the conductance from the fermionic perspective. At the fixed points well understood in the bosonic approach, such as the perfect transmission fixed point in the two-lead junction and the chiral fixed points in the Y-junction, the agreement of these results in both approaches is a necessary validation of the fermionic approach. On the other hand, for the fixed points eluding the bosonic treatment, such as the maximally open fixed point of the Y-junction (known as the ``$M$ fixed point''),\cite{1742-5468-2006-02-P02008,PhysRevB.68.035421,PhysRevLett.94.136405,PhysRevB.84.155426,PhysRevB.86.035137,PhysRevB.88.075131} these results can be directly compared to numerics\cite{PhysRevB.85.045120} where available.

In this work, we adopt the RPA fermionic approach to study the conductance tensor for a generic multi-lead junction in the presence of TLL leads. Our theory makes extensive use of the scattering basis transformation of the non-interacting part of the system; as a result it is explicitly formulated on the basis of the single-particle S-matrix (much like that in Ref.~\onlinecite{PhysRevB.66.165327}), and is formally independent of the number of wires. This stands in contrast to previous RPA treatments of junctions attached to FL leads, whose formulation is based on the renormalization of the conductance tensor instead and depends on the parametrization of the conductance tensor, different for two-lead junctions\cite{0295-5075-82-2-27001,PhysRevB.80.045109,LithJPhys.52.2353} and Y-junctions.\cite{PhysRevB.84.155426,PhysRevB.88.075131} We derive a Landauer-type conductance formula, appropriate for the renormalized S-matrix, and recover the additional contribution from the TLL leads to the conductance, absent in the naive Landauer formalism. Our theory is applied to the two-lead junction and Y-junction problems, where in addition to verifying existing results on the fixed points and the phase diagrams, the conductance of the $M$ fixed point attached to TLL leads is calculated. We summarize our findings below.

\begin{figure}
\includegraphics[width=0.6\textwidth]{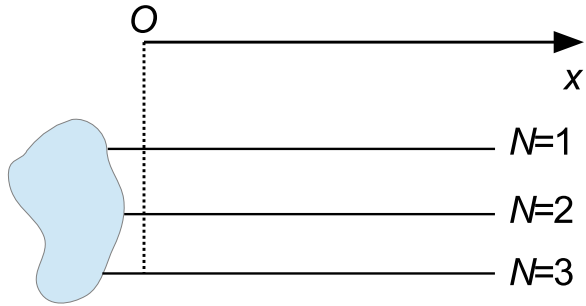}%
\caption{Sketch of the system with the number of leads $N=3$.\label{fig:sketch}}
\end{figure}

The system consists of $N$ quantum wires of interacting spinless electrons,
numbered $j=1$, $2$, ..., $N$, meeting at a junction which we choose as the
origin $x=0$. We align the wires so that they are parallel to the $+x$ axis;
see Fig.~\ref{fig:sketch}.

We assume that in the absence of interactions the junction is characterized by a
single-particle S-matrix, $S_{jj^{\prime}}$, independent of the energy of the
incident/scattered electron.  This is expected to be valid at low energies in the case of a non-resonant $S$-matrix, which we assume.
We adopt a Tomonaga-Luttinger model for the
electron-electron interaction in wire $j$,%
\begin{equation}
H_{\text{int}}^{j}=\int_{0}^{\infty}dxg_{2}^{j}\left(  x\right)  \psi
_{jR}^{\dag}\left(  x\right)  \psi_{jR}\left(  x\right)  \psi_{jL}^{\dag
}\left(  x\right)  \psi_{jL}\left(  x\right)  \text{,} \label{interaction0}%
\end{equation}
where $\psi_{jL/R}$ are low-energy left- and right-moving electrons in wire
$j$, and $g_{2}^{j}\left(  x\rightarrow\infty\right)  $ is a constant. A
finite $g_{2}^{j}\left(  \infty\right)  \neq0$ corresponds to a TLL lead
attached to wire $j$, while if $g_{2}^{j}\left(  \infty\right)  =0$ the
junction is considered to be connected to an FL lead. We define a dimensionless
interaction strength

\begin{equation}
\alpha_{j}\left(  x\right)  =g_{2}^{j}\left(  x\right)  /\left(  2\pi
v_{Fj}\right)  \text{,} \label{dimlessinteraction}%
\end{equation}
where $v_{Fj}$ is the Fermi velocity in wire $j$ without interaction.

The linear DC conductance tensor $G_{jj^{\prime}}$\ of the junction is defined
by $I_{j}=\sum_{j^{\prime}}G_{jj^{\prime}}V_{j^{\prime}}$, where $I_{j}$ is
the current flowing away from the junction in wire $j$, and $V_{j^{\prime}}$
is the bias voltage applied on lead $j^{\prime}$. In the absence of
interactions, $G_{jj^{\prime}}$ is given by the Landauer formula%

\begin{equation}
G_{jj^{\prime}}^{\text{FL}}=\frac{e^{2}}{2\pi}\left(  \delta_{jj^{\prime}%
}-\left\vert S_{jj^{\prime}}\right\vert ^{2}\right)  \text{.}
\label{Landauerintro}%
\end{equation}

In the first order perturbation theory in $\alpha_{j}$, where the
infrared singular corrections to physical observables are not resummed using
RG methods, attaching a junction to TLL leads as compared to FL leads changes
its linear DC conductance by%

\begin{equation}
G_{jj^{\prime}}^{\text{TLL}}-G_{jj^{\prime}}^{\text{FL}}=-\frac{e^{2}}{2\pi
}\sum_{n}\frac{1}{2}\alpha_{n}\left(  \infty\right)  \left(  \delta
_{jn}-\left\vert S_{jn}\right\vert ^{2}\right)  \left(  \delta_{nj^{\prime}%
}-\left\vert S_{nj^{\prime}}\right\vert ^{2}\right)  \text{.}
\label{OalphaTLLintro}%
\end{equation}

In the first order RG-improved perturbation theory, the bare S-matrix will be
replaced by the renormalized S-matrix at temperature $T$ in Eqs.~(\ref{Landauerintro}) and (\ref{OalphaTLLintro}).

In the RPA ``bare'' perturbation theory, the linear DC conductance tensors of the
same junction attached to TLL leads and FL leads are related by%
\begin{equation}
\mathbf{G}^{\text{TLL}}=\left(  \mathbf{1}-\mathbf{G}^{\text{FL}}%
\mathbf{G}_{c}^{-1}\right)  ^{-1}\mathbf{G}^{\text{FL}}\text{,}
\label{RPATLL1intro}%
\end{equation}
where $\mathbf{1}$\ is the $N\times N$ identity matrix, and $\mathbf{G}%
_{c}^{-1}$ is the contact resistance tensor between the wires and leads,%

\begin{equation}
\left(  G_{c}^{-1}\right)  _{jj^{\prime}}=\left(  \frac{e^{2}}{2\pi}\right)
^{-1}\frac{1}{2}\left[  1-\left(  K_{j}^{\text{L}}\right)  ^{-1}\right]
\delta_{jj^{\prime}}\text{.} \label{contresintro}%
\end{equation}
Here the bulk Luttinger parameter of the lead $j$ is given by $K_{j}%
^{\text{L}}=\sqrt{\left(  1-\alpha_{j}\left(  \infty\right)  \right)  /\left(
1+\alpha_{j}\left(  \infty\right)  \right)  }$. In the RPA RG-improved perturbation theory, the bare S-matrix is again
replaced by the renormalized S-matrix at temperature $T$ in Eqs.~(\ref{Landauerintro}) and (\ref{RPATLL1intro}).

For a $Z_{3}$ symmetric Y-junction at the maximally open $M$ fixed point,
$\left\vert S_{jj^{\prime}}\right\vert ^{2}=4/9-\delta_{jj^{\prime}}/3$. When
the junction is attached to TLL leads with dimensionless interaction strength
$\alpha$ and Luttinger parameter $K^{\text{L}}$, at the first order the
conductance tensor is%

\begin{equation}
G_{jj^{\prime}}^{\text{TLL,}M}=\left(  \frac{4}{9}-\frac{8}{27}\alpha\right)
\frac{e^{2}}{2\pi}\left(  3\delta_{jj^{\prime}}-1\right)  \text{.}
\label{OalphaMcondintro}%
\end{equation}
In the RPA, the conductance at $M$ becomes%

\begin{equation}
G_{jj^{\prime}}^{\text{TLL,}M}=\frac{4K^{\text{L}}}{3K^{\text{L}}+6}%
\frac{e^{2}}{2\pi}\left(  3\delta_{jj^{\prime}}-1\right)  \text{.}
\label{RPAMcondintro}%
\end{equation}
Also, the temperature dependence of the conductance with LL leads, as dictated by Eq. (\ref{RPATLL1intro}), is governed by the same fixed point exponents as those for FL leads at temperatures above the inverse lengths of interacting wires. For instance, near the $Z_{3}$ symmetric $M$ fixed point, in the presence of a quadratic perturbation at the junction which preserves both $Z_{3}$ symmetry and time-reversal symmetry, the correction to Eq. (\ref{OalphaMcondintro}) [or Eq. (\ref{RPAMcondintro})] scales as $T^{-2\alpha}$ [or $T^{-6\left( 1-K \right) / \left( 2+K \right) }$]; on the other hand, for a quadratic perturbation at the junction which preserves $Z_{3}$ symmetry but breaks time-reversal symmetry, the correction to Eq. (\ref{OalphaMcondintro}) [or Eq. (\ref{RPAMcondintro})] scales as $T^{\alpha /3}$ [or $T^{3\left( 1-K \right) \left( 2-K \right) / \left( 2+K \right) ^{2}}$].\cite{PhysRevB.66.165327,PhysRevB.84.155426,PhysRevB.88.075131} Whether a power law describes leading high-temperature or low-temperature behavior depends on the stability of $M$ with respect to that perturbation.

Eqs.~(\ref{OalphaTLLintro})--(\ref{RPAMcondintro}) are the main results of
this paper.

The rest of this paper is organized as follows. Section~\ref{sec:OalphaKubo} elaborates on our model for a generic multi-lead junction, and calculates its linear DC conductance to the first order in interaction. Section~\ref{sec:OalphaRG} is based on perturbative RG, again to the first order in interaction. We derive the
S-matrix RG equation in a Callan-Symanzik (CS) approach\cite{PhysRevB.80.045109} using the Kubo conductance calculated in Section~\ref{subsec:OalphaKubosub}. This establishes a modified Landauer formula involving the renormalized S-matrix in the case of FL leads. An additional contribution to the conductance, Eq.~(\ref{OalphaTLLintro}), is shown to arise from TLL leads. In Section~\ref{sec:RPA}, the conductance is found in the RPA to arbitrary order in interaction; we derive an S-matrix RG equation in the RPA, and again find the conductance [Eq.~(\ref{RPATLL1intro})]. Section~\ref{sec:offres}
applies our results to the fixed points of 2-lead junctions and Y-junctions at the first order and in the RPA. In particular, we find the conductance at the $M$ fixed point of a $Z_{3}$ symmetric Y-junction attached to TLL leads, Eqs.~(\ref{OalphaMcondintro}) and (\ref{RPAMcondintro}). Open questions are discussed in Section~\ref{sec:openquestions}. In Appendix~\ref{sec:appOalphaPT} we show details of the conductance calculations up to the first order in interaction. The Wilsonian derivation of the RG equation for the S-matrix\cite{PhysRevB.49.1966} is reviewed in
Appendix~\ref{sec:appOalphaNG}. Finally, the RPA conductance calculations are explained in Appendix~\ref{sec:appRPA}.

\section{First order perturbation theory of Kubo conductance\label{sec:OalphaKubo}}

In this section, we establish the model Hamiltonian, and present our results
for the linear DC conductance at the first order in interaction.

\subsection{Formulation of the problem\label{subsec:formulation}}

The system is modeled by a Hamiltonian consisting of three parts:%

\begin{equation}
H=\sum_{j=1}^{N}\left(  H_{0\text{,wire}}^{j}+H_{\text{int}}^{j}\right)
+H_{0,B}\text{.}%
\end{equation}
$H_{0\text{,wire}}^{j}$ is the non-interacting part of the Hamiltonian for
wire $j$, quadratic in electron operators, while the quartic $H_{\text{int}%
}^{j}$ term of Eq. (\ref{interaction0}) describes the electron-electron interaction in wire $j$. The
boundary term $H_{0,B}$ is quadratic, and is responsible for electron
transfer between wires across the junction. For simplicity we assume that each
wire only supports one single channel, and ignore quartic interactions between
wires, at the junction and between the junction and the wires.

In the continuum limit of the model, on each quantum wire we retain right- and
left-movers in narrow bands of wave vectors around the Fermi points $\pm
k_{Fj}$:%

\begin{equation}
\psi_{j}\left(  x\right)  \approx e^{ik_{Fj}x}\psi_{jR}\left(  x\right)
+e^{-ik_{Fj}x}\psi_{jL}\left(  x\right)  =\int_{-D}^{D}\frac{dE}{\sqrt{2\pi
v_{Fj}}}\left[  \psi_{jR}\left(  E\right)  e^{i\left(  \frac{E}{v_{Fj}}%
+k_{Fj}\right)  x}+\psi_{jL}\left(  E\right)  e^{-i\left(  \frac{E}{v_{Fj}%
}+k_{Fj}\right)  x}\right]  \text{,} \label{leftright}%
\end{equation}
where $v_{Fj}$ is the Fermi velocity in wire $j$, the dispersion relation is
$E=E_{j}\left(  k\right)  =v_{Fj}k$, and $D\ll v_{Fj}k_{Fj}$ is the
high-energy cutoff. Left-movers $\psi_{jL}$ are incident on the junction,
scattered, and turned into right-movers $\psi_{j^{\prime}R}$; $\psi_{jL}$ and
$\psi_{j^{\prime}R}$ are not independent degrees of freedom, but related by
the S-matrix of the junction [see also Eq.~(\ref{scatbas})]. The quadratic
part of the wire Hamiltonian now reads%

\begin{equation}
H_{0\text{,wire}}^{j}\approx iv_{Fj}\int_{0}^{\infty}dx\left[  \psi_{jL}%
^{\dag}\partial_{x}\psi_{jL}-\psi_{jR}^{\dag}\partial_{x}\psi_{jR}\right]
\left(  x\right)  \approx\int_{-D}^{D}dE\,E\left(  \psi_{jR}^{\dag}\left(
E\right)  \psi_{jR}\left(  E\right)  -\psi_{jL}^{\dag}\left(  E\right)
\psi_{jL}\left(  E\right)  \right)  \text{,} \label{bulkquad}%
\end{equation}

To model the electron-electron interaction, we assume it is short-ranged and
the system is away from half-filling, so that the Umklapp processes are
unimportant. We further ignore processes where two chiral densities of the
same chirality interact with one another, $\psi_{R}^{\dag}\psi_{R}\psi
_{R}^{\dag}\psi_{R}$ or $\psi_{L}^{\dag}\psi_{L}\psi_{L}^{\dag}\psi_{L}$;
these $g_{4}$\ processes\cite{solyom2010fundamentals3} renormalize the Fermi
velocity but do not change the Luttinger parameter by themselves. For spinless
fermions, this leaves us with only processes involving two chiral densities of
different chiralities, or $g_{2}$ processes, $\psi_{R}^{\dag}\psi_{R}\psi
_{L}^{\dag}\psi_{L}$. The electron-electron interaction is then represented by
a spatially variant $g_{2}$ term as in Eq.~(\ref{interaction0}).

\begin{figure}
\includegraphics[width=0.5\textwidth]{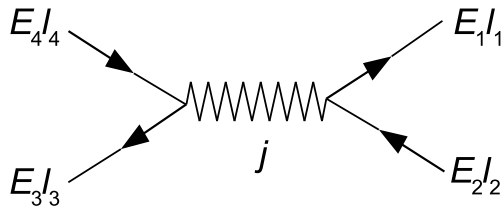}%
\caption{Diagrammatic representation of the electron-electron interaction.\label{fig:interaction}}
\end{figure}

Along the lines of Ref.~\onlinecite{PhysRevLett.71.3351,*PhysRevB.49.1966}, viewing the electron-electron interaction as a perturbation, we can first diagonalize the quadratic part of the Hamiltonian. The resultant eigenstates, which form the so-called scattering basis, can be related to the S-matrix in the low-energy theory. Note that such a scattering basis transformation is independent of the actual eigenstates of the fully interacting system; we are therefore always able to proceed with this transformation, regardless of whether the interaction is present at $x \to \infty$. For non-resonant scattering, which we assume throughout this paper, the S-matrix elements $S_{jj^{\prime}}\left(  E\right)  \equiv S_{jj^{\prime}}$ are independent of the electronic energy $E$, and the single-particle scattering state incident from wire $j^{\prime}$ with energy $E^{\prime}$ reads%

\begin{equation}
\phi_{j^{\prime}}^{\dag}\left(  E^{\prime}\right)  \left\vert 0\right\rangle
=\sum_{j}\int_{0}^{\infty}dx\frac{1}{\sqrt{2\pi v_{Fj}}}\left(  \delta
_{jj^{\prime}}e^{-i\frac{E^{\prime}}{v_{Fj}}x}\psi_{jL}^{\dag}\left(
x\right)  +S_{jj^{\prime}}e^{i\frac{E^{\prime}}{v_{Fj}}x}\psi_{jR}^{\dag
}\left(  x\right)  \right)  \left\vert 0\right\rangle +\cdots\text{,}
\label{scatstate}%
\end{equation}
where $\left\vert 0\right\rangle $ corresponds to  the filled Dirac sea, and the omitted terms
represent contributions from the junction area. Inverting Eq.~(\ref{scatstate}) we may express the original electrons $\psi$ in terms of the
scattering basis operators $\phi$,%

\begin{subequations}
\label{scatbas}%
\begin{align}
\psi_{jR}\left(  E\right)   &  =\sum_{j^{\prime}=1}^{N}\int dE^{\prime}%
\int_{0}^{\infty}dx\left(  \frac{1}{\sqrt{2\pi v_{Fj}}}e^{i\frac{E}{v_{Fj}}%
x}\right)  ^{\ast}\left(  \frac{1}{\sqrt{2\pi v_{Fj}}}S_{jj^{\prime}}%
e^{i\frac{E^{\prime}}{v_{Fj}}x}\right)  \phi_{j^{\prime}}\left(  E^{\prime
}\right) \nonumber\\
&  =\sum_{j^{\prime}}\int_{-D}^{D}\frac{dE^{\prime}}{2\pi}\frac{-i}%
{E-E^{\prime}-i0}S_{jj^{\prime}}\phi_{j^{\prime}}\left(  E^{\prime}\right)
\end{align}
Similarly%

\begin{equation}
\psi_{jL}\left(  E\right)  =\int_{-D}^{D}\frac{dE^{\prime}%
}{2\pi}\frac{i}{E-E^{\prime}+i0}\phi_{j}\left(  E^{\prime}\right)
\end{equation}
\end{subequations}
Now recast the Hamiltonian in the scattering basis. By definition, the
quadratic part of the Hamiltonian is diagonal:%

\begin{equation}
\sum_{j=1}^{N}H_{0\text{,wire}}^{j}+H_{0,B}=\sum_{j}\int dE\,E\phi_{j}^{\dag
}\left(  E\right)  \phi_{j}\left(  E\right)  \label{quadratic}%
\end{equation}
We insert the scattering basis transformation into the interaction Eq.~(\ref{interaction0}). Allowing the energies to run freely from $-\infty$ to
$\infty$ and calculating the energy integrals using the method of
residues,\cite{SovPhysJETP.38.202} we find%

\begin{equation}
H_{\text{int}}^{j}=\int_{0}^{\infty}dxg_{2}^{j}\left(  x\right)  \sum
_{l_{1}l_{2}l_{3}l_{4}}\int\frac{dE_{1}dE_{2}dE_{3}dE_{4}}{\left(
2\pi\right)  ^{2}v_{Fj}^{2}}\phi_{l_{1}}^{\dag}\left(  E_{1}\right)
\phi_{l_{2}}\left(  E_{2}\right)  \phi_{l_{3}}^{\dag}\left(  E_{3}\right)
\phi_{l_{4}}\left(  E_{4}\right)  e^{i\left(  -E_{1}+E_{2}+E_{3}-E_{4}\right)
\frac{x}{v_{Fj}}}S_{jl_{1}}^{\ast}S_{jl_{2}}\delta_{jl_{3}}\delta_{jl_{4}}%
\end{equation}
This is a plausible manipulation, seeing that the scattering basis
transformation should not introduce additional singularities at the band edge. Now%

\begin{equation}
H_{\text{int}}^{j}=\int_{0}^{\infty}dxg_{2}^{j}\left(  x\right)  \sum
_{l_{1}l_{2}l_{3}l_{4}}\int\frac{dE_{1}dE_{2}dE_{3}dE_{4}}{\left(
2\pi\right)  ^{2}v_{Fj}^{2}}\varrho_{l_{1}l_{2}l_{3}l_{4}}^{j}\left(
E_{1},E_{2},E_{3},E_{4};x\right)  \phi_{l_{1}}^{\dag}\left(  E_{1}\right)
\phi_{l_{2}}\left(  E_{2}\right)  \phi_{l_{3}}^{\dagger}\left(  E_{3}\right)
\phi_{l_{4}}\left(  E_{4}\right)  \text{,} \label{interaction}%
\end{equation}
where we introduce the function%

\begin{equation}
\varrho_{l_{1}l_{2}l_{3}l_{4}}^{j}\left(  E_{1},E_{2},E_{3},E_{4};x\right)
\equiv\frac{1}{2}\left[  e^{i\left(  -E_{1}+E_{2}+E_{3}-E_{4}\right)  \frac
{x}{v_{Fj}}}S_{jl_{1}}^{\ast}S_{jl_{2}}\delta_{jl_{3}}\delta_{jl_{4}%
}+e^{i\left(  -E_{3}+E_{4}+E_{1}-E_{2}\right)  \frac{x}{v_{Fj}}}S_{jl_{3}%
}^{\ast}S_{jl_{4}}\delta_{jl_{1}}\delta_{jl_{2}}\right]  \text{.}%
\end{equation}
Note that we have symmetrized the function $\varrho$ so that $\varrho
_{l_{1}l_{2}l_{3}l_{4}}^{j}\left(  E_{1},E_{2},E_{3},E_{4};x\right)
=\varrho_{l_{3}l_{4}l_{1}l_{2}}^{j}\left(  E_{3},E_{4},E_{1},E_{2};x\right)
$. This interaction is diagrammatically represented by the symmetric vertex in
Fig.~\ref{fig:interaction}. We may well opt not to symmetrize $\varrho$;
however, the two created electrons $E_{1}l_{1}$ and $E_{3}l_{3}$ (or the two
annihilated electrons $E_{2}l_{2}$ and $E_{4}l_{4}$) would be inequivalent in
that case, and the diagrammatic bookkeeping would be more difficult.

\subsection{Kubo conductance\label{subsec:OalphaKubosub}}

We now compute the linear DC conductance in Kubo formalism. The current
operator at coordinate $x$ in wire $j$ is first written in terms of the
fermion fields:%

\begin{equation}
\hat{I}_{j}\left(  x\right)  =ev_{Fj}\left(  \psi_{jR}^{\dag}\psi_{jR}%
-\psi_{jL}^{\dag}\psi_{jL}\right)  \left(  x\right)  \text{.}%
\end{equation}
Note that $\hat{I}_{j}$ is not changed by the interaction; it is proportional
to the commutator of the electron density with the Hamiltonian, but the
interaction commutes with the electron density. Using Eq.~(\ref{scatbas}) we
find the imaginary time correlation function $\Omega_{jj^{\prime}}\left(
x,x^{\prime};\tau-\tau^{\prime}\right)  \equiv-\left\langle T_{\tau}%
I_{j}\left(  x,\tau\right)  I_{j^{\prime}}\left(  x^{\prime},\tau^{\prime
}\right)  \right\rangle $ to be%

\begin{align}
&  \Omega_{jj^{\prime}}\left(  x,x^{\prime};\tau-\tau^{\prime}\right)
\nonumber\\
&  =-\frac{e^{2}}{\left(  2\pi\right)  ^{2}}\sum_{j_{1}j_{2}j_{1}^{\prime
}j_{2}^{\prime}}\int d\epsilon_{1}d\epsilon_{2}d\epsilon_{1}^{\prime}%
d\epsilon_{2}^{\prime}\left[  e^{i\frac{\epsilon_{2}-\epsilon_{1}}{v_{Fj}}%
x}S_{jj_{1}}^{\ast}S_{jj_{2}}-e^{-i\frac{\epsilon_{2}-\epsilon_{1}}{v_{Fj}}%
x}\delta_{jj_{1}}\delta_{jj_{2}}\right] \nonumber\\
&  \times\left[  e^{i\frac{\epsilon_{2}^{\prime}-\epsilon_{1}^{\prime}%
}{v_{Fj^{\prime}}}x^{\prime}}S_{j^{\prime}j_{1}^{\prime}}^{\ast}S_{j^{\prime
}j_{2}^{\prime}}-e^{-i\frac{\epsilon_{2}^{\prime}-\epsilon_{1}^{\prime}%
}{v_{Fj^{\prime}}}x^{\prime}}\delta_{j^{\prime}j_{1}^{\prime}}\delta
_{j^{\prime}j_{2}^{\prime}}\right]  \left\langle T_{\tau}\phi_{j_{1}}^{\dag
}\left(  \epsilon_{1},\tau\right)  \phi_{j_{2}}\left(  \epsilon_{2}%
,\tau\right)  \phi_{j_{1}^{\prime}}^{\dag}\left(  \epsilon_{1}^{\prime}%
,\tau^{\prime}\right)  \phi_{j_{2}^{\prime}}\left(  \epsilon_{2}^{\prime}%
,\tau^{\prime}\right)  \right\rangle _{\text{H}}\text{.} \label{curcorr}%
\end{align}
The imaginary time-ordered expectation value should be evaluated in the
Heisenberg picture. The linear DC conductance $G_{jj^{\prime}}$ is then given
by the retarded current-current correlation function $\Omega$,%

\begin{equation}
G_{jj^{\prime}}\left(  x,x^{\prime}\right)  =\lim_{\omega\rightarrow0}%
\lim_{\eta_{\omega}\rightarrow0^{+}}\frac{i}{\omega}\left[  \Omega
_{jj^{\prime}}\left(  x,x^{\prime};\omega^{+}\right)  -\Omega_{jj^{\prime}%
}\left(  x,x^{\prime};0\right)  \right]  \text{,} \label{Kubo}%
\end{equation}
where again $\omega^{+}\equiv\omega+i\eta_{\omega}$. The coordinate dependence
should vanish in the $\omega\rightarrow0$ limit, since where exactly we apply
the bias or measure the current is inconsequential in a DC
experiment.\cite{PhysRevB.54.R5239,1742-5468-2006-02-P02008}

Eq.~(\ref{Kubo}) is now calculated in perturbation theory. Switching to the interaction picture, we perform a Wick decomposition of the time-ordered product, go to the frequency space and sum over the Matsubara frequencies. The retarded correlation function is then obtained by analytic continuation $i\omega_{n}\rightarrow\omega^{+}\equiv\omega+i\eta_{\omega}$ where the $\eta_{\omega}\rightarrow0^{+}$ limit is taken. The energy integrals are calculated afterwards, followed by real space integrals [which appear in Eq.~(\ref{interaction})] in the end. Some details of this mostly standard calculation are given in Appendix~\ref{sec:appOalphaPT}; here again we only show the final results.

\begin{figure}
\includegraphics[width=0.9\textwidth]{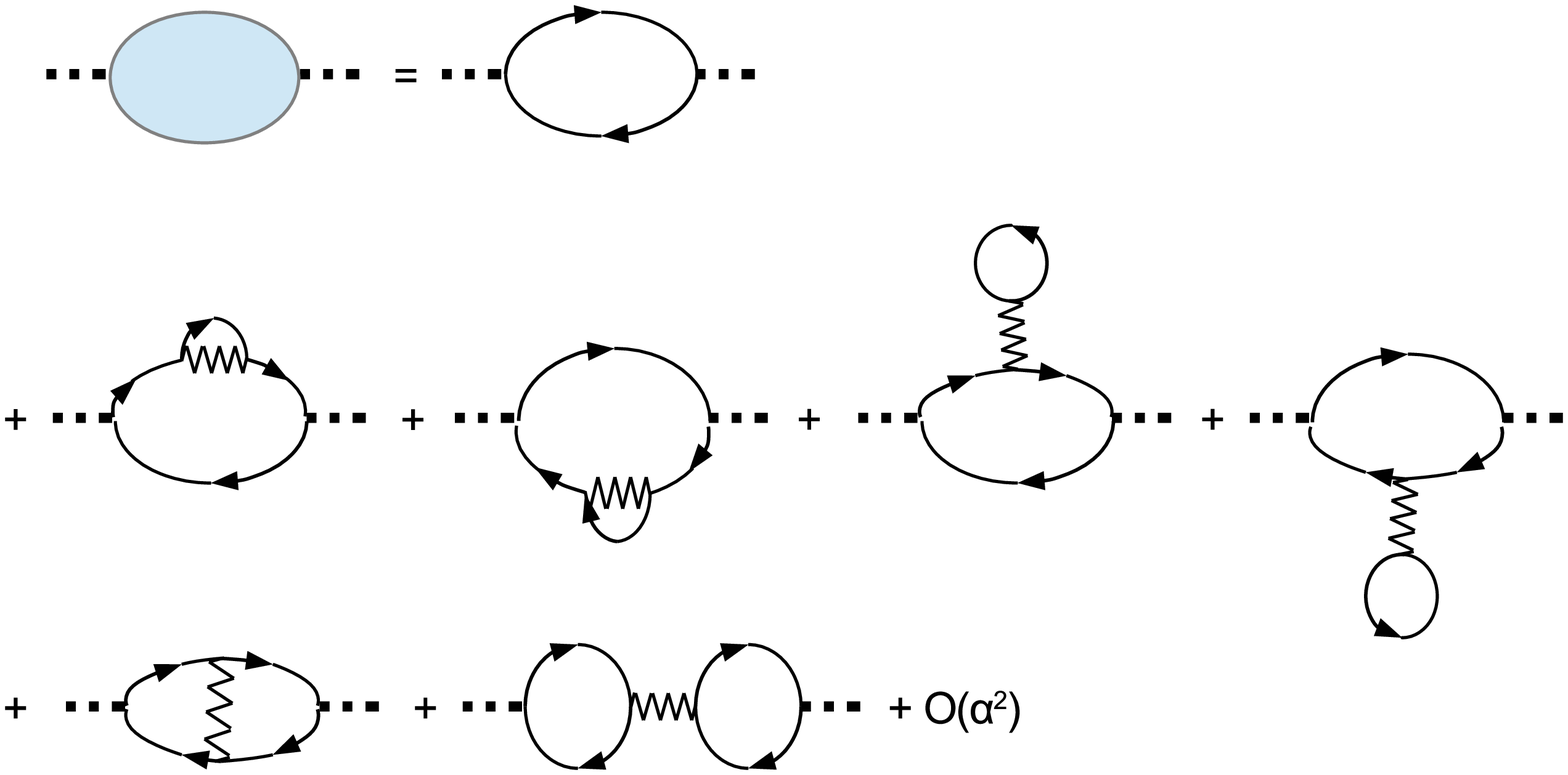}%
\caption{Diagrams contributing to the linear DC conductance at the first order in interaction. The second line shows the self-energy dressed bubble diagrams, while vertex correction diagrams are in the third line.\label{fig:Oalpha}}
\end{figure}

Feynman diagrams involved in the first order are shown in
Fig.~\ref{fig:Oalpha}. In the absence of interaction, we have a single bubble
diagram which leads to the usual linearized Landauer
formula:\cite{PhysRevLett.46.618,*PhysRevB.23.6851}%

\begin{equation}
G_{jj^{\prime}}^{\left(  0\right)  }=\frac{e^{2}}{2\pi}\left(  \delta
_{jj^{\prime}}-\left\vert S_{jj^{\prime}}\right\vert ^{2}\right)
\label{bubble}%
\end{equation}
Higher order diagrams can be classified into two basic types, namely
self-energy diagrams and vertex corrections. At the first order, contributions
from self-energy diagrams can be integrated into a Landauer-type formula:%

\begin{equation}
G_{jj^{\prime}}^{\left(  0\right)  }+G_{jj^{\prime}}^{\left(  1\right)
\text{,SE}}=\frac{e^{2}}{2\pi}\left(  \delta_{jj^{\prime}}-\int d\epsilon
\left[  -n_{F}^{\prime}\left(  \epsilon\right)  \right]  \left\vert
S_{jj^{\prime}}^{\text{d}\left(  1\right)  }\left(  \epsilon\right)
\right\vert ^{2}\right)  \label{OalphaFL}%
\end{equation}
where the first order ``dressed S-matrix'' $S^{\text{d}\left(  1\right)  }$\ is
given by%

\begin{align}
S_{jj^{\prime}}^{\text{d}\left(  1\right)  }\left(  \epsilon\right)   &
=S_{jj^{\prime}}-i\sum_{n}\int_{0}^{\infty}dy\,\alpha_{n}\left(  y\right)
\int\frac{d\epsilon^{\prime}}{v_{Fn}}n_{F}\left(  \epsilon^{\prime}\right)
\nonumber\\
&  \times S_{jn}S_{nn}^{\ast}S_{nj^{\prime}}\exp\left(  2i\left(
\epsilon-\epsilon^{\prime}\right)  \frac{y}{v_{Fn}}\right)  +\delta_{jn}%
S_{nn}\delta_{j^{\prime}n}\exp\left(  -2i\left(  \epsilon-\epsilon^{\prime
}\right)  \frac{y}{v_{Fn}}\right)  \text{;} \label{OalphaSdress0}%
\end{align}
$n_{F}\left(  \epsilon\right)  =1/\left(  e^{\beta\epsilon}+1\right)  $ is the
Fermi distribution at temperature $\beta=1/T$, and $\alpha_{n}\left(
y\right)  $ is defined in Eq.~(\ref{dimlessinteraction}). For a
non-interacting system $S_{jj^{\prime}}^{\text{d}\left(  1\right)  }\left(
\epsilon\right)  =S_{jj^{\prime}}$; this is in agreement with our intuitive expectation.

We now perform the $y$ integral in a simple model. Let us assume that the
junction is connected through wire $n$ to a TLL or FL lead at $x=L_{n}$; in
other words, when $x\geq L_{n}$, $\alpha_{n}\left(  x\right)  =\alpha
_{n}\left(  \infty\right)  $ becomes a constant independent of $x$ and
$d\alpha_{n}\left(  x\right)  /dx=0$. We further assume that the interaction
inside the wire is also uniform, i.e.%

\begin{equation}
\alpha_{j}\left(  x\right)  =\alpha_{j}\left(  0\right)  +\left[  \alpha
_{j}\left(  \infty\right)  -\alpha_{j}\left(  0\right)  \right]  \theta\left(
x-L_{j}\right)  \label{interactionmodel}%
\end{equation}
where $\theta\left(  x\right)  $ is the Heaviside unit-step function.
Integrating over $y$:%

\begin{align}
S_{jj^{\prime}}^{\text{d}\left(  1\right)  }\left(  \epsilon\right)   &
=S_{jj^{\prime}}\left(  \epsilon\right)  -\sum_{n}\int d\epsilon^{\prime}%
\frac{n_{F}\left(  \epsilon^{\prime}\right)  }{2\left(  \epsilon^{\prime
}-\epsilon\right)  }\left(  S_{jn}S_{nn}^{\ast}S_{nj^{\prime}}\left[  \left(
\alpha_{n}\left(  \infty\right)  -\alpha_{n}\left(  0\right)  \right)
e^{2i\left(  \epsilon-\epsilon^{\prime}\right)  \frac{L_{n}}{v_{Fn}}}%
+\alpha_{n}\left(  0\right)  \right]  \right. \nonumber\\
&  \left.  -\delta_{jn}S_{nn}\delta_{j^{\prime}n}\left[  \left(  \alpha
_{n}\left(  \infty\right)  -\alpha_{n}\left(  0\right)  \right)  e^{-2i\left(
\epsilon-\epsilon^{\prime}\right)  \frac{L_{n}}{v_{Fn}}}+\alpha_{n}\left(
0\right)  \right]  \right)  \text{.} \label{OalphaSdress}%
\end{align}
The $\epsilon^{\prime}$ integral is infrared divergent, which prompts an RG
resummation of leading logarithms. We will determine the renormalization of
the S-matrix using Eq.~(\ref{OalphaSdress}) and discuss its implications in
Section~\ref{sec:OalphaRG}.

The vertex corrections, in the meanwhile, contribute a completely different type of terms:%

\begin{align}
&  G_{jj^{\prime}}^{\left(  1\right)  \text{,VC}}\left(  x,x^{\prime}\right)
=-\frac{e^{2}}{2\pi}i\sum_{n}\int_{0}^{\infty}\frac{dy}{v_{Fn}}\alpha
_{n}\left(  y\right) \nonumber\\
&  \times\lim_{\omega\rightarrow0}\lim_{\eta_{\omega}\rightarrow0^{+}}%
\omega\left[  \left\vert S_{jn}\right\vert ^{2}\delta_{j^{\prime}n}%
e^{i\omega^{+}\left(  \frac{x}{v_{Fj}}-\frac{x^{\prime}}{v_{Fj^{\prime}}%
}\right)  }e^{2i\omega^{+}\frac{y}{v_{Fn}}}\theta\left(  \frac{y}{v_{Fn}%
}-\frac{x^{\prime}}{v_{Fj^{\prime}}}\right)  \right. \nonumber\\
&  \left.  -\delta_{jn}\delta_{j^{\prime}n}e^{-i\omega^{+}\left(  \frac
{x}{v_{Fj}}+\frac{x^{\prime}}{v_{Fj^{\prime}}}\right)  }e^{2i\omega^{+}%
\frac{y}{v_{Fn}}}\theta\left(  \frac{y}{v_{Fn}}-\frac{x}{v_{Fj}}\right)
\theta\left(  \frac{y}{v_{Fn}}-\frac{x^{\prime}}{v_{Fj^{\prime}}}\right)
\right. \nonumber\\
&  \left.  -\delta_{jn}\delta_{j^{\prime}n}e^{i\omega^{+}\left(  \frac
{x}{v_{Fj}}+\frac{x^{\prime}}{v_{Fj^{\prime}}}\right)  }e^{-2i\omega^{+}%
\frac{y}{v_{Fn}}}\theta\left(  \frac{x}{v_{Fj}}-\frac{y}{v_{Fn}}\right)
\theta\left(  \frac{x^{\prime}}{v_{Fj^{\prime}}}-\frac{y}{v_{Fn}}\right)
\right. \nonumber\\
&  \left.  -\left\vert S_{jn}\right\vert ^{2}\left\vert S_{nj^{\prime}%
}\right\vert ^{2}e^{i\omega^{+}\left(  \frac{x}{v_{Fj}}+\frac{x^{\prime}%
}{v_{Fj^{\prime}}}\right)  }e^{2i\omega^{+}\frac{y}{v_{Fn}}}\right.
\nonumber\\
&  \left.  +\delta_{jn}\left\vert S_{nj^{\prime}}\right\vert ^{2}%
e^{-i\omega^{+}\left(  \frac{x}{v_{Fj}}+\frac{x^{\prime}}{v_{Fj^{\prime}}%
}\right)  }e^{2i\omega^{+}\frac{y}{v_{Fn}}}\theta\left(  \frac{y}{v_{Fn}%
}-\frac{x}{v_{Fj}}\right)  \right]  \label{OalphaTLL0}%
\end{align}
Integration by parts gives us%

\begin{equation}
\frac{2i\omega^{+}}{v_{Fn}}\int_{y_{l}}^{y_{u}}dy\alpha_{n}\left(  y\right)
e^{2i\omega^{+}\frac{y}{v_{Fn}}}=\alpha_{n}\left(  y_{u}\right)
e^{2i\omega^{+}\frac{y_{u}}{v_{Fn}}}-\alpha_{n}\left(  y_{l}\right)
e^{2i\omega^{+}\frac{y_{l}}{v_{Fn}}}-\int_{y_{l}}^{y_{u}}dy\,e^{2i\omega
^{+}\frac{y}{v_{Fn}}}\frac{d\alpha_{n}\left(  y\right)  }{dy}\text{,}
\label{g2yint}%
\end{equation}
where $y_{u}$ can be $v_{Fn}x/v_{Fj}$,\ $v_{Fn}x^{\prime}/v_{Fj^{\prime}}$\ or
$\infty$, and $y_{l}$ can be $v_{Fn}x/v_{Fj}$,\ $v_{Fn}x^{\prime
}/v_{Fj^{\prime}}$ or $0$. We can let $x$ and $x^{\prime}$ be sufficiently
large so that $y_{u}>L_{n}$ is always satisfied; thus in the $d\alpha_{n}/dy$
term in Eq.~(\ref{g2yint}), $y_{u}$ may be replaced by $L_{n}$.

If $y_{u}\rightarrow\infty$, the $\alpha_{n}\left(  y_{u}\right)  $ term damps
out due to the small imaginary part $\eta_{\omega}$, and Eq.~(\ref{g2yint})
becomes in the $\omega\rightarrow0$ and $\eta_{\omega}\rightarrow0$ limit%

\begin{subequations}
\begin{equation}
\frac{2i\omega^{+}}{v_{Fn}}\int_{y_{l}}^{y_{u}}dy\alpha_{n}\left(  y\right)
e^{2i\omega^{+}\frac{y}{v_{Fn}}}=-\alpha_{n}\left(  y_{l}\right)  -\int
_{y_{l}}^{L_{n}}dy\frac{d\alpha_{n}\left(  y\right)  }{dy}=-\alpha_{n}\left(
L_{n}\right)  =-\alpha_{n}\left(  \infty\right)  \text{.}%
\end{equation}
On the other hand, if $y_{u}$ is finite, the $\alpha_{n}\left(  y_{u}\right)
$ term will survive the $\omega\rightarrow0$ and $\eta_{\omega}\rightarrow0$ limit:%

\begin{equation}
\frac{2i\omega^{+}}{v_{Fn}}\int_{y_{l}}^{y_{u}}dy\alpha_{n}\left(  y\right)
e^{2i\omega^{+}\frac{y}{v_{Fn}}}=\alpha_{n}\left(  y_{u}\right)  -\alpha
_{n}\left(  y_{l}\right)  -\int_{y_{l}}^{L_{n}}dy\frac{d\alpha_{n}\left(
y\right)  }{dy}=\alpha_{n}\left(  y_{u}\right)  -\alpha_{n}\left(
L_{n}\right)  =0\text{.}%
\end{equation}
\end{subequations}
Therefore, taking the DC limit explicitly in Eq.~(\ref{OalphaTLL0}), we find
wire $n$ contributes to the vertex correction only when it is attached to a
TLL lead, and the interaction inside the wire is immaterial:%

\begin{equation}
G_{jj^{\prime}}^{\left(  1\right)  \text{,VC}}\left(  x,x^{\prime}\right)
=-\frac{e^{2}}{2\pi}\sum_{n}\frac{1}{2}\alpha_{n}\left(  \infty\right)
\left(  \delta_{jn}-\left\vert S_{jn}\right\vert ^{2}\right)  \left(
\delta_{nj^{\prime}}-\left\vert S_{nj^{\prime}}\right\vert ^{2}\right)
\text{.} \label{OalphaTLL}%
\end{equation}
When $\alpha_{n}\left(  \infty\right)  =0$, as is the case for any wire $n$
attached to an FL lead, the vertex correction due to $n$ vanishes.

\section{First-order Callan-Symanzik perturbative RG\label{sec:OalphaRG}}

In this section, we analyze the result of Section~\ref{sec:OalphaKubo} from
the perspective of the CS formulation of RG,\cite{PhysRevB.80.045109} and
present a modified Landauer formula involving the renormalized S-matrix in the
case of FL leads, supplemented by vertex corrections from TLL leads.

Intuitively, once the renormalization flow of the S-matrix is stopped by a physical infrared cutoff, the renormalized S-matrix should represent the non-interacting part of the low-energy theory of the junction, and can be taken as an input to the Landauer formalism. However, such an argument does not address the role of the low-energy residual interaction, which turns out to be especially important in the case of TLL leads. Also, in principle, the Landauer formalism is
well-founded only in the absence of inelastic scattering. We are therefore motivated to study the conductance in the CS formulation, which fully exposes possible deviations from the Landauer predictions.

In the CS formulation of RG, we start from a field theory with a running cutoff $D$, and
calculate low-energy physical observables (in our case the linear DC conductance tensor $G_{jj^{\prime}}$) as a function of the running coupling constants of the theory [in our case the S-matrix elements $S_{jj^{\prime}}\left(  D\right)  $]. This is once again accomplished by perturbation theory in interaction, in formal analogy to Section~\ref{sec:OalphaKubo}. However, the crucial difference is that we are now expressing certain low-energy physical quantities in terms of running coupling constants, whereas in Section~\ref{sec:OalphaKubo} we calculate the corresponding renormalized quantities in terms of bare coupling constants. We require that when $D$ is greater than the energy scales at which the system is
probed, namely the finite temperature $T$, $G_{jj^{\prime}}$ should be independent of $D$. Therefore, by allowing the cutoff to run from $D$ to $D-\delta D$, where $\delta D\ll D$, we can find the RG equation satisfied by the coupling constants $S_{jj^{\prime}}\left(  D\right)  $.

Beginning from the simplest case where all leads are FL leads, $\alpha
_{n}\left(  \infty\right)  =0$ for all $n$, the vertex correction Eq.~(\ref{OalphaTLL}) vanishes, and the full linear DC conductance to $O\left( \alpha\right)  $ is given by Eq.~(\ref{OalphaFL}). Reducing the cutoff from $D$ to $D-\delta D$ and demanding the right-hand side of Eq.~(\ref{OalphaFL}) be a scaling invariant, we have%

\begin{equation}
\int_{\delta D}d\epsilon\left[  -n_{F}^{\prime}\left(  \epsilon\right)
\right]  \left\vert S_{jj^{\prime}}^{\text{d}\left(  1\right)  }\left(
\epsilon,D\right)  \right\vert ^{2}+\int_{-D}^{D}d\epsilon\left[
-n_{F}^{\prime}\left(  \epsilon\right)  \right]  \left[  \left(
S_{jj^{\prime}}^{\text{d}\left(  1\right)  }\left(  \epsilon,D\right)
\right)  ^{\ast}\delta S_{jj^{\prime}}^{\text{d}\left(  1\right)  }\left(
\epsilon,D\right)  +\text{c.c.}\right]  =0 \label{OalphaFLCallanSymanzik}%
\end{equation}
where

\begin{equation}
\delta S_{jj^{\prime}}^{\text{d}\left(  1\right)  }\left(  \omega,D\right)
\equiv S_{jj^{\prime}}^{\text{d}\left(  1\right)  }\left(  \omega,D\right)
-S_{jj^{\prime}}^{\text{d}\left(  1\right)  }\left(  \omega,D-\delta D\right)
\text{,} \label{deltaOalphaSdress}%
\end{equation}
Here $S_{jj^{\prime}}^{\text{d}\left(  1\right)  }\left(  \epsilon,D\right)
$ is Eq.~(\ref{OalphaSdress}) with the $\epsilon^{\prime}$ integral going from
$-D$ to $D$, and all S-matrix elements understood to be cutoff-dependent,
$S_{jj^{\prime}}\rightarrow S_{jj^{\prime}}\left(  D\right)  $.

Since the derivative of the Fermi function is peaked at the Fermi energy with
width $T$, the $\int_{\delta D}$ integral in Eq.~(\ref{OalphaFLCallanSymanzik}) approximately vanishes while $D\gtrsim T$; Eq.~(\ref{OalphaFLCallanSymanzik}) is thus automatically satisfied if Eq.~(\ref{deltaOalphaSdress}) vanishes. The implication is that, at least in the case of FL leads, the renormalization of the conductance can be fully accounted for by the renormalization of the S-matrix.

To the lowest order in $\delta D$, the condition that $\delta S_{jj^{\prime}}^{\text{d}\left(  1\right)  }\left(  \omega,D\right) =0$ is equivalent to%

\begin{align}
\delta S_{jj^{\prime}}\left(  \omega,D\right)   &  \equiv S_{jj^{\prime}%
}\left(  \omega,D\right)  -S_{jj^{\prime}}\left(  \omega,D-\delta D\right)
\nonumber\\
&  =\sum_{n}\int_{\delta D}d\epsilon^{\prime}\frac{n_{F}\left(  \epsilon
^{\prime}\right)  }{2\left(  \epsilon^{\prime}-\omega\right)  }\left(
S_{jn}S_{nn}^{\ast}S_{nj^{\prime}}\left[  \left(  \alpha_{n}\left(
\infty\right)  -\alpha_{n}\left(  0\right)  \right)  e^{2i\left(
\omega-\epsilon^{\prime}\right)  \frac{L_{n}}{v_{Fn}}}+\alpha_{n}\left(
0\right)  \right]  \right. \nonumber\\
&  \left.  -\delta_{jn}S_{nn}\delta_{j^{\prime}n}\left[  \left(  \alpha
_{n}\left(  \infty\right)  -\alpha_{n}\left(  0\right)  \right)  e^{-2i\left(
\omega-\epsilon^{\prime}\right)  \frac{L_{n}}{v_{Fn}}}+\alpha_{n}\left(
0\right)  \right]  \right)  \text{.} \label{Srenor0}%
\end{align}
where $\int_{\delta D}=\int_{\left(  D-\delta D\right)  }^{D}+\int
_{-D}^{-\left(  D-\delta D\right)  }$ stands for integration over fast modes.

If $D\gtrsim\left\vert \omega\right\vert $, $\epsilon^{\prime}-\omega$ can be
approximated as $\pm2D$, thus giving rise to a scaling contribution $O\left(
\delta D/D\right)  $. If $D\gtrsim v_{Fn}/L_{n}$, $\exp\left(  \pm
i2DL_{n}/v_{Fn}\right)  $ oscillates rapidly with $D$ and is negligible; on
the other hand, when $D\lesssim v_{Fn}/L_{n}$, $\exp\left(  \pm i2DL_{n}%
/v_{Fn}\right)  \approx1$. Finally, if $D\gtrsim T$, the factors $n_{F}\left(
D\right)  \approx0$ and $n_{F}\left(  -D\right)  \approx1$ are approximately
independent of $D$. Therefore, to $O\left(  \delta D/D\right)  $, Eq.~(\ref{Srenor0}) predicts that%

\begin{equation}
\delta S_{jj^{\prime}}=-\frac{\delta D}{2D}\left(  \sum_{n}\alpha_{n}\left(
D\right)  S_{jn}S_{nn}^{\ast}S_{nj^{\prime}}-\alpha_{j}\left(  D\right)
S_{jj}\delta_{jj^{\prime}}\right)  \text{,} \label{Srenor}%
\end{equation}
independent of $\omega$, provided $D\gtrsim\max\left\{  \left\vert
\omega\right\vert ,T\right\}  $. Here we have defined a cutoff-dependent
interaction strength%

\begin{equation}
\alpha_{n}\left(  D\right)  \equiv%
\genfrac{\{}{.}{0pt}{}{\alpha_{n}\left(  0\right)  \text{, }D\gtrsim
v_{Fn}/L_{n}}{\alpha_{n}\left(  \infty\right)  \text{, }D\lesssim v_{Fn}%
/L_{n}}%
\text{.} \label{alphaD}%
\end{equation}
This means the renormalization will stop at the energy scale of the
incident/scattered electron or the temperature, whichever is higher. In
addition, the energy scale associated with the inverse length of wire $n$,
$v_{Fn}/L_{n}$, determines whether the renormalization due to that wire is
controlled by interaction strength in the wire $\alpha_{n}\left(  0\right)  $
or that in the lead $\alpha_{n}\left(  \infty\right)  $: the effective
interaction strength crosses over from $\alpha_{n}\left(  0\right)  $ to
$\alpha_{n}\left(  \infty\right)  $ as the $D$ is reduced below $v_{Fn}/L_{n}$.

We are now in a position to write down the RG equation for the S-matrix valid
to $O\left(  \alpha\right)  $. Restoring the explicit cutoff dependence, we have%

\begin{equation}
-\frac{dS_{jj^{\prime}}\left(  D\right)  }{d\ln D}=-\frac{1}{2}\sum_{n}%
\alpha_{n}\left(  D\right)  \left[  S_{jn}\left(  D\right)  S_{nn}^{\ast
}\left(  D\right)  S_{nj^{\prime}}\left(  D\right)  -\delta_{j^{\prime}%
n}S_{nn}\left(  D\right)  \delta_{nj}\right]  \label{NazarovGlazman}%
\end{equation}
where the RG flow is cut off at the temperature $T$. This is the equation
given in Refs.~\onlinecite{PhysRevLett.71.3351,PhysRevB.66.165327}. It can
be readily checked that Eq.~(\ref{NazarovGlazman}) preserves the unitarity of
the S-matrix.

We pause to remark that, as the cutoff is reduced below the inverse length of
one of the wires, renormalization due to that wire is governed only by the
lead to which that wire is attached. This is reasonable because a junction of
finite-length TLL wires attached to FL leads should, at low energies,
renormalize into a junction connected directly to FL
leads.\cite{PhysRevB.54.R5239,1742-5468-2006-02-P02008,PhysRevB.80.045109}

Returning to the conductance analysis, once the cutoff $D$ is
reduced to the order of $T$, the perturbative correction to the S-matrix
$S_{jj^{\prime}}^{\text{d}\left(  1\right)  }\left(  \epsilon,D\right)
-S_{jj^{\prime}}\left(  D\right)  $ vanishes to the scaling accuracy; thus
$S_{jj^{\prime}}\left(  D=T\right)  $ may be used to approximate the dressed
S-matrix in Eq.~(\ref{OalphaFL}), and the conductance for a junction connected
to FL leads is given by the modified Landauer formula,%

\begin{equation}
G_{jj^{\prime}}^{\text{FL}}=\frac{e^{2}}{2\pi}\left(  \delta_{jj^{\prime}%
}-\left\vert S_{jj^{\prime}}\left(  T\right)  \right\vert ^{2}\right)
\text{,} \label{OalphaFL1}%
\end{equation}
where the S-matrix is now fully renormalized according to Eq.~(\ref{NazarovGlazman}), with the cutoff reduced to the temperature $T$. This
is the Landauer-type formula invoked in Refs.~\onlinecite{PhysRevLett.71.3351,*PhysRevB.49.1966,PhysRevB.66.165327}.

When some of the leads are TLL leads, corrections of Eq.~(\ref{OalphaTLL})
must also be taken into account. It is important to note, however, that in a
CS analysis of the total conductance, Eq.~(\ref{deltaOalphaSdress}) remains
valid to $O\left(  \alpha\right)  $. This is because as the cutoff is lowered,
Eq.~(\ref{OalphaTLL}) contributes additional terms of the form of
$\alpha\left(  \infty\right)  S^{\ast}\delta S$ to Eq.~(\ref{OalphaFLCallanSymanzik}). However, by Eq.~(\ref{deltaOalphaSdress}),
$\delta S$ is $O\left(  \alpha\right)  $; hence $\alpha\left(  \infty\right)
S^{\ast}\delta S$ is $O\left(  \alpha^{2}\right)  $, and is negligible to
$O\left(  \alpha\right)  $.

To calculate the total conductance at $D=T$ with TLL leads, we go slightly
beyond the first order and dress the $O\left(  \alpha\right)  $ vertex
correction diagrams with $O\left(  \alpha\right)  $ self-energy diagrams,
shown in Fig.~\ref{fig:OalphaVC}. The bare S-matrix in Eq.~(\ref{OalphaTLL})
is then replaced by the dressed S-matrix, $S^{\text{d}\left(  1\right)  }$:%

\begin{figure}
\includegraphics[width=0.7\textwidth]{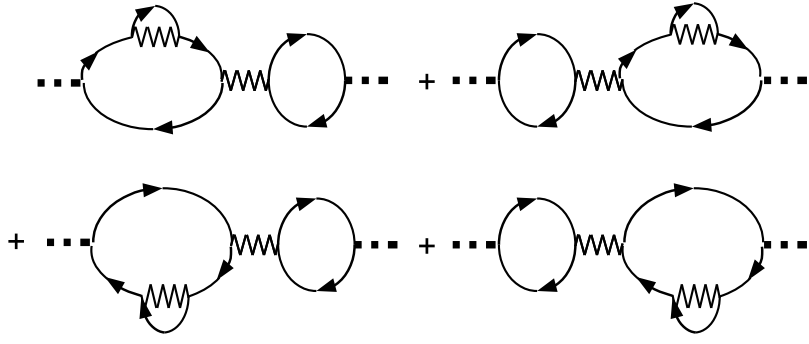}%
\caption{Dressing of the first order vertex correction diagrams by the first order self-energy diagrams.\label{fig:OalphaVC}}
\end{figure}

\begin{align}
G_{jj^{\prime}}^{\text{d}\left(  1\right)  \text{,VC}}\left(  x,x^{\prime
}\right)   &  =-\frac{e^{2}}{2\pi}\sum_{n}\frac{1}{2}\alpha_{n}\left(
\infty\right)  \left(  \delta_{jn}-\int d\epsilon_{1}\left[  -n_{F}^{\prime
}\left(  \epsilon_{1}\right)  \right]  \left\vert S_{jn}^{\text{d}\left(
1\right)  }\left(  \epsilon_{1}\right)  \right\vert ^{2}\right) \nonumber\\
&  \times\left(  \delta_{nj^{\prime}}-\int d\epsilon_{2}\left[  -n_{F}%
^{\prime}\left(  \epsilon_{2}\right)  \right]  \left\vert S_{nj^{\prime}%
}^{\text{d}\left(  1\right)  }\left(  \epsilon_{2}\right)  \right\vert
^{2}\right)  \text{.}%
\end{align}
This allows us to repeat our previous analysis for the case of FL leads, and
further approximate $S_{jj^{\prime}}^{\text{d}\left(  1\right)  }$ by
$S_{jj^{\prime}}\left(  D=T\right)  $. Thus the TLL leads contribute an
additional conductance of%

\begin{equation}
G_{jj^{\prime}}^{\text{TLL}}-G_{jj^{\prime}}^{\text{FL}}=-\frac{e^{2}}{2\pi
}\sum\nolimits_{n}\frac{\alpha_{n}\left(  \infty\right)  }{2}\left(
\delta_{jn}-\left\vert S_{jn}\left(  T\right)  \right\vert ^{2}\right)
\left(  \delta_{nj^{\prime}}-\left\vert S_{nj^{\prime}}\left(  T\right)
\right\vert ^{2}\right)  \text{.} \label{OalphaTLL1}%
\end{equation}

Eqs.~(\ref{NazarovGlazman}), (\ref{OalphaFL1}) and (\ref{OalphaTLL1}) provide
a comprehensive first-order picture for non-resonant tunneling through a
junction: the interaction renormalizes the S-matrix, the renormalized S-matrix
determines the conductance through a Landauer-type formula if the junction is
connected to FL leads, and the residual interaction further modifies the
conductance if the junction is attached to TLL leads. As will be demonstrated
in Section~\ref{sec:RPA}, this picture is by no means limited to the first order.

\section{S-matrix renormalization and conductance in the RPA}

\label{sec:RPA}

\begin{figure}
\includegraphics[width=0.7\textwidth]{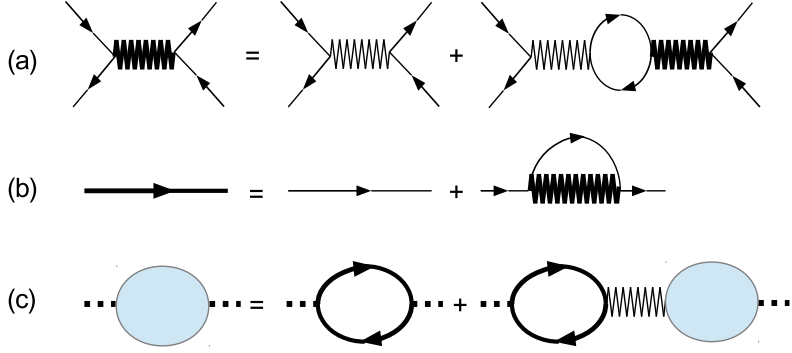}
\caption{The RPA diagrammatics: (a) effective interaction in the RPA represented by thick wavy lines; (b) dressed propagator in the RPA, to $O \left ( \delta D/D \right )$ in RG, represented by thick straight lines; and (c) diagrams contributing to the Kubo conductance in the RPA. The dressed propagator in (b) is calculated to $O \left ( \delta D/D \right )$ only, because higher order terms in $\delta D/D$ do not contribute to the renormalization of the S-matrix [Eq.~(\ref{RPANazarovGlazman})]--- see Section~\ref{sec:OalphaRG} for an explanation in the first order context. (a) and (c) do not involve truncation at $O \left ( \delta D/D \right )$ because any renormalization of the interaction [Eq.~(\ref{RPAvertex})] and the conductance [Eqs.~(\ref{RPAFL}) and (\ref{RPATLL})] can be attributed to the renormalization of the S-matrix. Note that (c) features a thin interaction line (rather than a thick one) to avoid double-counting.\label{fig:RPA}}
\end{figure}

In this section, we extends our first-order RG analysis in Section~\ref{sec:OalphaRG} to arbitrary order in interaction under the
RPA.\cite{0295-5075-82-2-27001,PhysRevB.80.045109,LithJPhys.52.2353,PhysRevB.84.155426,PhysRevB.88.075131} The correlation function Eq.~(\ref{curcorr}) is perturbatively evaluated for
both self-energy diagrams and vertex corrections by the same procedures,
except that the interaction is dressed with ring diagrams; see
Fig.~\ref{fig:RPA}. We subsequently find the S-matrix RG equation in the CS
scheme and express the conductance in terms of the renormalized S-matrix. This is once more a
straightforward calculation, and we simply present the outcome, leaving the
details for Appendix~\ref{sec:appRPA}.

Introduce the shorthand $W_{jj^{\prime}}\left(  D\right)  \equiv\left\vert
S_{jj^{\prime}}\left(  D\right)  \right\vert ^{2}$. The RPA self-energy
diagrams give rise to a modified Landauer formula:%

\begin{equation}
G_{jj^{\prime}}^{\text{FL}}=\frac{e^{2}}{2\pi}\left[  \delta_{jj^{\prime}%
}-W_{jj^{\prime}}\left(  T\right)  \right]  \text{,} \label{RPAFL}%
\end{equation}
where the renormalization of the S-matrix is governed by a generalization of
Eq.~(\ref{NazarovGlazman}),%

\begin{equation}
-\frac{dS_{jj^{\prime}}\left(  D\right)  }{d\ln D}=-\frac{1}{2}\sum
_{n_{1}n_{2}}\left[  S_{jn_{1}}\left(  D\right)  \Pi_{n_{1}n_{2}}\left(
D\right)  S_{n_{2}n_{1}}^{\ast}\left(  D\right)  S_{n_{2}j^{\prime}}\left(
D\right)  -\delta_{j^{\prime}n_{1}}\Pi_{n_{1}n_{2}}^{\ast}\left(  D\right)
S_{n_{2}n_{1}}\left(  D\right)  \delta_{n_{2}j}\right]  \text{.}
\label{RPANazarovGlazman}%
\end{equation}
The RPA-dressed interaction is%

\begin{equation}
\Pi\left(  D\right)  \equiv2\left[  \mathcal{Q}\left(  D\right)  -W\left(
D\right)  \right]  ^{-1}\text{,} \label{RPAvertex}%
\end{equation}
where%

\begin{equation}
\mathcal{Q}_{jj^{\prime}}=Q_{j}\left(  D\right)  \delta_{jj^{\prime}}\text{,
}Q_{j}\left(  D\right)  =\frac{1+K_{j}\left(  D\right)  }{1-K_{j}\left(
D\right)  }\text{,} \label{scriptQ}%
\end{equation}
with $K_{j}\left(  D\right)  =\sqrt{\left(  1-\alpha_{j}\left(  D\right)
\right)  /\left(  1+\alpha_{j}\left(  D\right)  \right)  }$ being the
cutoff-dependent ``Luttinger parameter'' for wire $j$; $\alpha_{j}\left(
D\right)  $ is given in Eq.~(\ref{alphaD}). To lowest order in $\alpha_j$, $\Pi_{ij}=\delta_{ij}\alpha_{j}$. When all wires of the junction are attached to FL leads, in parallel with the $O\left(  \alpha\right)  $
calculation, Eq.~(\ref{RPAFL}) captures the entirety of the conductance. This
is in agreement with the Kubo formula calculation in Refs.~\onlinecite{0295-5075-82-2-27001,PhysRevB.80.045109,LithJPhys.52.2353,PhysRevB.84.155426,PhysRevB.88.075131}
in the language of chiral fermion densities.

When some wires are attached to TLL leads, they again provide important
corrections to the DC conductance. All RPA vertex correction diagrams dressed
with RPA self-energy evaluate to%

\begin{equation}
G_{jj^{\prime}}^{\text{TLL}}-G_{jj^{\prime}}^{\text{FL}}=-\frac{e^{2}}{2\pi
}\sum\nolimits_{n_{1}n_{2}}\left[  \delta_{jn_{1}}-W_{jn_{1}}\left(  T\right)
\right]  \frac{1}{2}\Pi_{n_{1}n_{2}}^{\text{L}}\left[  \delta_{n_{2}j^{\prime
}}-W_{n_{2}j^{\prime}}\left(  T\right)  \right]  \text{,} \label{RPATLL}%
\end{equation}
where the residual effective interaction is%

\begin{equation}
\Pi^{\text{L}}=2\left[  \mathcal{Q}^{\text{L}}-W\left(  T\right)  \right]
^{-1}\text{,}%
\end{equation}
and $\mathcal{Q}^{\text{L}}$ is given by Eq.~(\ref{scriptQ}) with $K_{j}$
replaced by $K_{j}^{\text{L}}=\sqrt{\left(  1-\alpha_{j}\left(  \infty\right)
\right)  /\left(  1+\alpha_{j}\left(  \infty\right)  \right)  }$, the
Luttinger parameter of the lead.

Remarkably, if we define the DC contact resistance tensor between the wires
and leads,%

\begin{equation}
\left(  G_{c}^{-1}\right)  _{jj^{\prime}}=\left(  \frac{e^{2}}{2\pi}\right)
^{-1}\frac{1}{2}\left[  1-\left(  K_{j}^{\text{L}}\right)  ^{-1}\right]
\delta_{jj^{\prime}}\text{,}%
\end{equation}
then Eq.~(\ref{RPATLL}) can be formally recast as%

\begin{equation}
\mathbf{G}^{\text{TLL}}=\left(  \mathbf{1}-\mathbf{G}^{\text{FL}}%
\mathbf{G}_{c}^{-1}\right)  ^{-1}\mathbf{G}^{\text{FL}}\text{,}
\label{RPATLL1}%
\end{equation}
where $\mathbf{1}$\ is the $N\times N$ identity matrix. The same relation has
been derived in Refs.~\onlinecite{1742-5468-2006-02-P02008,PhysRevB.86.075451}, which assume that
the DC contact resistance between a finite TLL wire and an FL lead is not
affected by the junction at the other end of the TLL wire. This intuitive
assumption is reinforced by our calculations.

We emphasize that the inclusion of the vertex correction diagrams does not change the RG equation of the S-matrix, Eq.~(\ref{RPANazarovGlazman}). [The TLL leads do change the renormalization of the S-matrix through the scale-dependent interaction, Eq.~(\ref{alphaD}).] The reason for this is as follows. Eq.~(\ref{RPANazarovGlazman}) results from dressing the single particle propagator as shown in panel (b) of Fig.~\ref{fig:RPA}. The conductance is calculated in perturbation theory by replacing all bare single particle propagators (the thin lines) with the dressed ones (the thick lines) in the basic bubble diagram and the vertex correction diagrams; or equivalently, by replacing all bare S-matrix elements with the ones dressed with the RPA self-energy. As with the case at the first order, the RPA vertex correction diagrams do not introduce additional cutoff-sensitive integrals, and all cutoff-sensitive integrals originate from the dressed S-matrix. Therefore, the dressed S-matrix should be a cutoff-independent quantity when we apply the CS scheme to the conductance, regardless of whether the vertex correction diagrams contribute to the conductance. The form of Eq.~(\ref{RPANazarovGlazman}) is thus independent of vertex corrections.

An immediate consequence of the robustness of the S-matrix renormalization is that, in the temperature range above the inverse lengths of the wires, the universal scaling exponents of the conductance versus temperature are the same for FL leads and TLL leads to the accuracy of the RG method. Since the temperature dependence of the residual interaction ultimately results from that of the renormalized $W$ matrix, Taylor-expanding Eq. (\ref{RPATLL}) in the vicinity of a fixed point, we find that the scaling exponents of the TLL lead conductance are none other than those of the $W$ matrix, i.e. those of the FL lead conductance; in this temperature range the TLL leads merely modify the non-universal multiplicative coefficients to the power law. Therefore, for the temperature dependence of the conductance in Section~\ref{sec:intro}, we have directly quoted the FL lead results from Refs.~\onlinecite{PhysRevB.66.165327,PhysRevB.84.155426,PhysRevB.88.075131}. We also note that Eq. (\ref{RPATLL1}) is always valid in the temperature range above the inverse lengths of the wires, which allows us to determine in the RPA the full temperature dependence of the TLL lead conductance using that of the FL lead conductance. The full crossover of FL lead conductance has been worked out analytically for two-lead junctions\cite{0295-5075-82-2-27001,PhysRevB.80.045109,LithJPhys.52.2353} and for certain special cases of Y-junctions.\cite{PhysRevB.84.155426}

Eqs.~(\ref{RPAFL})--(\ref{RPATLL1}) are the central results of this paper.
They show that at least in the RPA, in addition to the Landauer-type formula,
TLL leads give rise to important corrections to the linear DC conductance
which are also given in terms of the renormalized S-matrix. In the remainder
of this paper, we implement these results in non-resonant tunneling through
2-lead junctions and Y-junctions.

\section{Fixed point conductance\label{sec:offres}}

In this section we evaluate the conductance at several established fixed points of 2-lead junctions and Y-junctions attached to FL leads and TLL leads. The analysis is carried out at the first order in interaction [Eqs.~(\ref{OalphaFL1}) and (\ref{OalphaTLL1})] and then in the RPA [Eqs.~(\ref{RPAFL}) and (\ref{RPATLL1})]. In particular, we will examine the conductance of the maximally open $M$ fixed point in the RPA for the $Z_{3}$ symmetric Y-junction.

For simplicity, the interactions are once more modeled by Eq.~(\ref{interactionmodel}). We write $\alpha_{j}\left(  0\right)  $, the
interaction strength in wire $j$, simply\ as $\alpha_{j}$; also when the
junction is connected to TLL leads, we assume the interactions in wires and
leads are uniform and identical, i.e. $\alpha_{j}\left(  \infty\right)
=\alpha_{j}$. Of course, by definition $\alpha_{j}\left(  \infty\right)  =0$
for FL leads.

\subsubsection{2-lead junction}

In a 2-lead junction of spinless fermions away from resonance, solving the S-matrix RG equations [Eq.~(\ref{NazarovGlazman}) at the first order and Eq.~(\ref{RPANazarovGlazman}) in the RPA], we find that the only fixed points are the complete reflection fixed point [the $N$ (Neumann) fixed point] and the perfect transmission fixed point [the $D$ (Dirichlet) fixed point].\cite{PhysRevLett.71.3351,*PhysRevB.49.1966,0295-5075-82-2-27001,PhysRevB.80.045109,LithJPhys.52.2353}

At the $N$ fixed point $W_{12}=0$, the two wires are decoupled from each
other, and we find the obvious result that the conductance $G_{jj^{\prime}%
}^{N\text{,FL}}=G_{jj^{\prime}}^{N\text{,TLL}}=0$, irrespective of what leads
the junction is attached to.

On the other hand, at the $D$ fixed point $W_{12}=1$, the backscattering
between the two wires vanishes. With FL leads $G_{jj^{\prime}}^{D\text{,FL}%
}=\left(  e^{2}/2\pi\right)  \left(  2\delta_{jj^{\prime}}-1\right)  $, as
predicted by the naive Landauer formula; with TLL leads, Eq.~(\ref{OalphaTLL1}) predicts

\begin{equation}
G_{jj^{\prime}}^{D\text{,TLL}}=\left(  1-\frac{\alpha_{1}+\alpha_{2}}%
{2}\right)  \frac{e^{2}}{2\pi}\left(  2\delta_{jj^{\prime}}-1\right)%
\end{equation}
at the first order, and Eq.~(\ref{RPATLL1}) predicts

\begin{equation}
G_{jj^{\prime}}^{D\text{,TLL}}=\frac{2K_{1}K_{2}}{K_{1}+K_{2}}\frac{e^{2}%
}{2\pi}\left(  2\delta_{jj^{\prime}}-1\right)%
\end{equation}
in the RPA. Here the RPA has recovered the famous result for the conductance of two semi-infinite TLL
wires.\cite{PhysRevB.52.R17040}%

\subsubsection{Y-junction}

Even at the first order in interaction, the RG flow portrait for a Y-junction is more complicated than the two-lead junction.\cite{PhysRevB.66.165327} Solving Eq.~(\ref{NazarovGlazman}), we find a ``non-geometrical'' $M$ fixed point whose existence and transmission probabilities generally depend on the interaction strengths, in addition to the ``geometrical'' fixed points $N$, $A_{j}$ and $\chi^{\pm}$. Provided the interactions are not too strong, these are also the only fixed points allowed in the RPA.\cite{PhysRevB.88.075131} $N$ (complete reflection) and $A_{j}$ (asymmetric) can be obtained by adding a third decoupled wire with label $j$ to the $N$ and $D$ fixed points of the two-lead junction respectively. The conductances at $N$ and $A_{j}$ are therefore a trivial generalization of the two-lead case, and we will focus on $\chi^{\pm}$ and $M$ alone.

At the chiral fixed points $\chi^{\pm}$,
in the absence of interaction, an electron incident from wire $j$ is perfectly
transmitted to wire $j\pm1$ (here we identify $j+3\equiv j$); thus the
time-reversal symmetry is broken. The $W$ matrix is given by $W_{jj^{\prime}%
}=\left(  1-\delta_{jj}\mp\epsilon_{jj^{\prime}}\right)  /2$, where the
anti-symmetric tensor $\epsilon_{jj^{\prime}}$ is defined by $\epsilon
_{12}=\epsilon_{23}=\epsilon_{31}=1$, $\epsilon_{21}=\epsilon_{32}%
=\epsilon_{13}=-1$ and $\epsilon_{jj}=0$. At the first order, inserting the $W$ matrix into Eqs.~(\ref{OalphaFL1})
and (\ref{OalphaTLL1}), we find $G_{jj^{\prime}}^{\chi^{\pm}\text{,FL}%
}=\left(  e^{2}/2\pi\right)  \left(  3\delta_{jj^{\prime}}-1\pm\epsilon
_{jj^{\prime}}\right)  /2$, and%

\begin{equation}
G_{jj^{\prime}}^{\chi^{\pm}\text{,TLL}}-G_{jj^{\prime}}^{\chi^{\pm}\text{,FL}%
}=-\frac{e^{2}}{2\pi}\frac{1}{2}\left[  \left(  \alpha_{j}+\alpha_{j^{\prime}%
}\right)  \left(  \frac{3}{2}-\delta_{jj^{\prime}}\right)  +\frac{1}{2}\left(
\alpha_{1}+\alpha_{2}+\alpha_{3}\right)  \left(  1-\delta_{jj^{\prime}}%
\pm\epsilon_{jj^{\prime}}\right)  \right]\text{.}%
\end{equation}
In the RPA, on the other hand, Eq.~(\ref{RPATLL1}) gives the conductance at $\chi^{\pm}$ with TLL leads as%

\begin{equation}
G_{jj^{\prime}}^{\chi^{\pm}\text{,TLL}}=2\frac{e^{2}}{2\pi}\frac{K_{j}\left(
K_{1}+K_{2}+K_{3}\right)  \delta_{jj^{\prime}}+\left(  \pm K_{1}K_{2}%
K_{3}\epsilon_{jj^{\prime}}-K_{j}K_{j^{\prime}}\right)  }{K_{1}+K_{2}%
+K_{3}+K_{1}K_{2}K_{3}}\text{,}%
\end{equation}
which agrees with the result of bosonization analysis.\cite{PhysRevB.86.075451}

The presence of the $M$ fixed point can be inferred
in a $Z_{3}$ symmetric time-reversal invariant Y-junction with attractive
interactions: in this system, $N$ is unstable, $A_{j}$ is forbidden by $Z_{3}$
symmetry, and $\chi^{\pm}$ are forbidden by time-reversal symmetry, so there
must be at least one stable fixed point. The $W$ matrix has
generally interaction-dependent elements at $M$. At the first order,%

\begin{equation}
W_{jj^{\prime}}=%
\genfrac{\{}{.}{0pt}{}{\left(  \frac{\alpha_{1}\alpha_{2}\alpha_{3}/\alpha
_{j}}{\alpha_{1}\alpha_{2}+\alpha_{2}\alpha_{3}+\alpha_{3}\alpha_{1}}\right)
^{2}\text{, }j=j^{\prime}}{\left(  1-\frac{\alpha_{1}\alpha_{2}\alpha
_{3}/\alpha_{j}}{\alpha_{1}\alpha_{2}+\alpha_{2}\alpha_{3}+\alpha_{3}%
\alpha_{1}}\right)  \left(  1-\frac{\alpha_{1}\alpha_{2}\alpha_{3}%
/\alpha_{j^{\prime}}}{\alpha_{1}\alpha_{2}+\alpha_{2}\alpha_{3}+\alpha
_{3}\alpha_{1}}\right)  \text{, }j\neq j^{\prime}}%
\text{.} \label{OalphaMW}%
\end{equation}
We see explicitly that $M$ obeys time-reversal symmetry, $W_{jj^{\prime}%
}=W_{j^{\prime}j}$. Demanding $0\leq W_{jj^{\prime}}\leq1$, we find that at the first-order $M$
can only exist in the following situations: 1) $\alpha_{1}$, $\alpha_{2}$,
$\alpha_{3}>0$; 2) $\alpha_{1}$, $\alpha_{2}$, $\alpha_{3}<0$; 3) $\alpha
_{1}>0$, $\alpha_{2}>0$, $\alpha_{3}<-\alpha_{1}$, $\alpha_{3}<-\alpha_{2}$;
4) $\alpha_{1}<0$, $\alpha_{2}<0$, $\alpha_{3}>-\alpha_{1}$, $\alpha
_{3}>-\alpha_{2}$; and situations equivalent to 3) and 4) up to permuted
subscripts (e.g. $\left(  \alpha_{1},\alpha_{2},\alpha_{3}\right)
\rightarrow\left(  \alpha_{3},\alpha_{1},\alpha_{2}\right)  $).

Substituting Eq.~(\ref{OalphaMW}) into Eq.~(\ref{OalphaTLL1}), we find that at the first-order the conductance at $M$ obeys%

\begin{equation}
G_{jj^{\prime}}^{M\text{,TLL}}-G_{jj^{\prime}}^{M\text{,FL}}=%
\genfrac{\{}{.}{0pt}{}{-\frac{e^{2}}{2\pi}\frac{\left(  \alpha_{1}+\alpha
_{2}\right)  \left(  \alpha_{2}+\alpha_{3}\right)  \left(  \alpha_{1}%
+\alpha_{3}\right)  }{2\left(  \alpha_{1}\alpha_{2}+\alpha_{2}\alpha
_{3}+\alpha_{3}\alpha_{1}\right)  ^{3}}\alpha_{j}^{2}\left(  \alpha_{1}%
+\alpha_{2}+\alpha_{3}-\alpha_{j}\right)  ^{2}\text{, }j=j^{\prime}%
}{\frac{e^{2}}{2\pi}\frac{\left(  \alpha_{1}+\alpha_{2}\right)  \left(
\alpha_{2}+\alpha_{3}\right)  \left(  \alpha_{1}+\alpha_{3}\right)  }{2\left(
\alpha_{1}\alpha_{2}+\alpha_{2}\alpha_{3}+\alpha_{3}\alpha_{1}\right)  ^{3}%
}\left[  \alpha_{j}\alpha_{j^{\prime}}\left(  \alpha_{1}\alpha_{2}+\alpha
_{2}\alpha_{3}+\alpha_{3}\alpha_{1}\right)  -\frac{\left(  \alpha_{1}%
\alpha_{2}\alpha_{3}\right)  ^{2}}{\alpha_{j}\alpha_{j^{\prime}}}\right]
\text{, }j\neq j^{\prime}}%
\text{.}%
\end{equation}

Note that for $Z_{3}$ symmetric interactions ($\alpha_{j}=\alpha$),
$W_{jj^{\prime}}=1/9+\delta_{jj^{\prime}}/3$ becomes independent of the
interaction strength. Now $W_{jj^{\prime}}$ produces the maximal transmission
probability $8/9$ allowed by unitarity in a $Z_{3}$ symmetric S-matrix, and at the first order
$G_{jj^{\prime}}^{M\text{,TLL}}-G_{jj^{\prime}}^{M\text{,FL}}=-\left(
8/27\right)  \alpha\left(  e^{2}/2\pi\right)  \left(  2\delta_{jj^{\prime}%
}-1\right)  $. Compared to FL leads, TLL leads enhance conductance for
attractive interactions and reduce conductance for repulsive interactions, as
with the two-lead $D$ fixed point.

In the RPA, the $W$ matrix of the $M$ fixed point is generally cumbersome, but reduces to the aforementioned maximally transmitting $W$ matrix for $Z_{3}$ symmetric interactions. Eq.~(\ref{RPATLL1}) then gives%

\begin{equation}
G_{jj^{\prime}}^{M\text{,TLL}}=\frac{4K}{3K+6}\frac{e^{2}}{2\pi}\left(
3\delta_{jj^{\prime}}-1\right)  \text{.}%
\end{equation}
This result supports the findings of Ref.~\onlinecite{PhysRevB.85.045120}.
There the $M$ fixed point conductance of a Y-junction of infinite TLL wires is
computed numerically using DMRG, and
conjectured to be%

\begin{equation}
G_{jj^{\prime}}=\frac{2K\gamma}{2K+3\gamma-3K\gamma}\frac{e^{2}}{2\pi}\left(
3\delta_{jj^{\prime}}-1\right)  \text{,}%
\end{equation}
where it is suggested that the dimensionless parameter $\gamma$ is $4/9$ based
on the non-interacting limit $K=1$.

\section{Discussion and open questions\label{sec:openquestions}}

In this paper, using the fermionic RG formalism, we calculated the linear DC conductance tensor of a junction of multiple quantum wires. We showed, both at the first order and in the RPA, that a junction attached to FL leads has a conductance tensor which obeys a linearized Landauer-type formula with a renormalized S-matrix. TLL leads modify the conductance through vertex corrections, and the conductance with FL leads may be heuristically related to the conductance with TLL leads through the contact resistance between leads. In this section, we would like to discuss some of the questions left open in our approach.

First, we have assumed that scattering by the junction is fully described by operators which are quadratic in conduction fermions and independent of other degrees of freedom. Local operators quartic in fermions are ignored, among others. This does not pose a threat to the first-order calculations, because any quartic local operator has a scaling dimension of at least $4\times 1/2=2$ in the non-interacting case, and is necessarily highly irrelevant. However, it has been shown that sufficiently strong attractive bulk interactions can render quartic boundary operators relevant.\cite{1742-5468-2006-02-P02008} An example is the electron pair hopping operator at the $Z_{3}$ symmetric Y-junction, $\psi_{1L}^{\dag}\psi_{1R}^{\dag}\psi_{3}\partial_{x}\psi_{3}\left( x=0\right) +\text{h.c.}$: it is of dimension $3/K$ at the asymmetric fixed point $A_{3}$, where $K$ is the Luttinger parameter of all three wires, and $A_{3}$ sees wire 3 decoupled from perfectly connected wires 1 and 2. Apparently, for very strong interactions $K>3$, this operator becomes relevant and can potentially dominate the physical properties of the stable fixed point. Unfortunately, the present RPA analysis does not predict a scaling exponent consistent with this operator;\cite{PhysRevB.88.075131} it is hence incomplete in this regard, and should not be carried too far into the regime of strongly attractive bulk interactions.

A related issue is the existence of the $D$ fixed points in the Y-junction. Predicted by the bosonic approaches\cite{PhysRevB.59.15694,1742-5468-2006-02-P02008,PhysRevB.86.075451} but not the fermionic ones,\cite{PhysRevB.84.155426,PhysRevB.88.075131} these fixed points are only stable for strong attractive interactions. They are most notably characterized by Andreev reflections, even when electron-electron interaction is absent in the bulk. This hints at multi-particle scattering at the junction, and rules out the possibility to represent the $D$ fixed points by single-particle S-matrices. (Single-particle S-matrices with particle-hole channels are not feasible either since the $D$ fixed points respect particle number conservation.)\cite{1742-5468-2006-02-P02008} The $D$ fixed points are not predicted by purely fermionic approaches, because the latter are based on the ansatz that the junction is always described by a single-particle S-matrix along the RG flow; but such an ansatz will likely be invalidated if, for instance, relevant quartic boundary operators are present. We are thus led to believe that the lack of $D$ fixed points in the present RPA analysis does not refute their possible stability when the bulk interactions are strongly attractive. Indeed, the refermionization method adopted by Ref.~\onlinecite{PhysRevB.92.125138} may be successfully used to describe the crossover from the ``pair tunneling'' $D$ fixed point to the $\chi^{\pm}$ fixed points in the vicinity of Luttinger parameter $K=3$, with an S-matrix of free fermions which are not the original electrons.

On the other hand, even when the bulk interactions are relatively weak, it is not a priori clear to what extent the RPA is successful. In the Tomonaga-Luttinger model (which we have adopted in our bulk quantum wires), the RPA is known to be exact due to the interaction which separately conserves the numbers of right- and left-movers.\cite{SovPhysJETP.38.202} This is no longer the case once right- and left-movers become mixed up by the scattering at the junction. It has been pointed out that going beyond the RPA changes the renormalization of the S-matrix away from the ``geometrical'' fixed points, although all universal scaling exponents stay the same.\cite{0295-5075-82-2-27001,PhysRevB.80.045109,PhysRevB.84.155426} As for the ``non-geometrical'' $M$ fixed point in the Y-junction, its position is generally shifted when we go beyond the RPA. Remarkably, however, if the interaction is $Z_{3}$ symmetric, not only the $W$ matrix but also the scaling exponents at the $M$ fixed point remain identical with the RPA results up to the third order in interaction.\cite{PhysRevB.84.155426} The agreement of our RPA result with the numerics of Ref.~\onlinecite{PhysRevB.85.045120} is suggestive, but more work on vertex corrections is required to verify the validity of our RPA conductance at the $Z_{3}$ symmetric $M$ fixed point with TLL leads, Eq.~(\ref{RPAMcondintro}).

\begin{acknowledgments}
This work was supported in part by NSERC of Canada, Discovery Grant 36318-2009 (ZS and IA) and the Canadian Institute for Advanced Research (IA). The authors would like to acknowledge helpful discussions with D. Giuliano, L. I. Glazman, Y. Komijani and A. Rahmani during the course of this work, and also V. Meden and D. G. Polyakov for bringing multiple important references to their attention. The authors gratefully acknowledge the hospitality of GGI Florence where part of this work was done.
\end{acknowledgments}

\appendix

\section{Details of zeroth and first order perturbation theory\label{sec:appOalphaPT}}

In this appendix we present some of the crucial steps in the perturbative
calculation of the conductance up to the first order
in interaction, which lead to Eqs.~(\ref{OalphaFL}) and (\ref{OalphaTLL}). We go through the standard procedures for the
conductance calculation at the zeroth order in interaction, then highlight the
treatments specific to the first order.

\subsection{Zeroth order\label{subsec:conductancezero}}

At the zeroth order, there is only one bubble diagram for the current-current
correlation function. Wick's theorem gives%

\begin{align}
&  \left\langle T_{\tau}\phi_{j_{1}}^{\dag}\left(  \epsilon_{1},\tau\right)
\phi_{j_{2}}\left(  \epsilon_{2},\tau\right)  \phi_{j_{1}^{\prime}}^{\dag
}\left(  \epsilon_{1}^{\prime},\tau^{\prime}\right)  \phi_{j_{2}^{\prime}%
}\left(  \epsilon_{2}^{\prime},\tau^{\prime}\right)  \right\rangle \nonumber\\
&  =-\delta_{j_{2}j_{1}^{\prime}}\delta\left(  \epsilon_{2}-\epsilon
_{1}^{\prime}\right)  \mathcal{G}_{j_{2}}\left(  \epsilon_{2},\tau
-\tau^{\prime}\right)  \delta_{j_{1}j_{2}^{\prime}}\delta\left(  \epsilon
_{1}-\epsilon_{2}^{\prime}\right)  \mathcal{G}_{j_{1}}\left(  \epsilon
_{1},\tau^{\prime}-\tau\right)  \text{.}%
\end{align}
Here $\mathcal{G}$ is the free scattering basis Matsubara Green's function
$\mathcal{G}_{j}\left(  E,i\omega_{n}\right)  =1/\left(  i\omega_{n}-E\right)
$, $\omega_{n}=\left(  2n+1\right)  \pi/\beta$. We have dropped the H
subscript in Eq.~(\ref{curcorr}) when switching to the interaction picture.
Going to the frequency space, doing the standard Matsubara sum%

\begin{equation}
\frac{1}{\beta}\sum_{i\omega_{n}}\mathcal{G}_{j_{2}}\left(  E_{2},i\omega
_{n}\right)  \mathcal{G}_{j_{1}}\left(  E_{1},i\omega_{n}-ip_{m}\right)
=\frac{n_{F}\left(  E_{2}\right)  -n_{F}\left(  E_{1}\right)  }{-ip_{m}%
+E_{2}-E_{1}}\text{,} \label{bubbleMats}%
\end{equation}
where $p_{m}=2m\pi/\beta$ is a bosonic frequency, and performing analytic
continuation $ip_{m}\rightarrow\omega^{+}\equiv\omega+i\eta_{\omega}$
($\eta_{\omega}\rightarrow0^{+}$)\ yield the zeroth order retarded correlation function,%

\begin{align}
&  \Omega_{jj^{\prime}}^{\left(  0\right)  }\left(  x,x^{\prime};\omega
^{+}\right)  =\frac{e^{2}}{\left(  2\pi\right)  ^{2}}\sum_{j_{1}j_{2}}\int
d\epsilon_{1}d\epsilon_{2}\frac{n_{F}\left(  \epsilon_{2}\right)
-n_{F}\left(  \epsilon_{1}\right)  }{-\omega^{+}+\epsilon_{2}-\epsilon_{1}%
}\left[  \delta_{jj^{\prime}}e^{i\left(  \epsilon_{2}-\epsilon_{1}\right)
\left(  \frac{x}{v_{Fj}}-\frac{x^{\prime}}{v_{Fj^{\prime}}}\right)  }\right.
\nonumber\\
&  \left.  +\delta_{jj^{\prime}}e^{i\left(  \epsilon_{1}-\epsilon_{2}\right)
\left(  \frac{x}{v_{Fj}}-\frac{x^{\prime}}{v_{Fj^{\prime}}}\right)
}-\left\vert S_{j^{\prime}j}\right\vert ^{2}e^{i\left(  \epsilon_{1}%
-\epsilon_{2}\right)  \left(  \frac{x}{v_{Fj}}+\frac{x^{\prime}}%
{v_{Fj^{\prime}}}\right)  }-\left\vert S_{jj^{\prime}}\right\vert
^{2}e^{i\left(  \epsilon_{2}-\epsilon_{1}\right)  \left(  \frac{x}{v_{Fj}%
}+\frac{x^{\prime}}{v_{Fj^{\prime}}}\right)  }\right]  \text{.}
\label{zerothOmega1}%
\end{align}
We have done the $j_{1}$ and $j_{2}$ sums using unitarity of the S-matrix.
Employing contour techniques, we integrate over $\epsilon_{1}$ on $\left(
-\infty,\infty\right)  $ for the term proportional to $n_{F}\left(
\epsilon_{2}\right)  $, and integrate over $\epsilon_{2}$ on $\left(
-\infty,\infty\right)  $ for the term proportional to $n_{F}\left(
\epsilon_{1}\right)  $:%

\begin{align}
&  \Omega_{jj^{\prime}}^{\left(  0\right)  }\left(  x,x^{\prime};\omega
^{+}\right) \nonumber\\
&  =\frac{e^{2}}{\left(  2\pi\right)  ^{2}}\int d\epsilon_{2}\left(  2\pi
i\right)  n_{F}\left(  \epsilon_{2}\right)  \left[  \delta_{jj^{\prime}%
}e^{i\omega^{+}\left\vert \frac{x}{v_{Fj}}-\frac{x^{\prime}}{v_{Fj^{\prime}}%
}\right\vert }-0-\left\vert S_{jj^{\prime}}\right\vert ^{2}e^{i\omega
^{+}\left(  \frac{x}{v_{Fj}}+\frac{x^{\prime}}{v_{Fj^{\prime}}}\right)
}\right] \nonumber\\
&  -\frac{e^{2}}{\left(  2\pi\right)  ^{2}}\int d\epsilon_{1}\left(  2\pi
i\right)  n_{F}\left(  \epsilon_{1}\right)  \left[  \delta_{jj^{\prime}%
}e^{i\omega^{+}\left\vert \frac{x}{v_{Fj}}-\frac{x^{\prime}}{v_{Fj^{\prime}}%
}\right\vert }-0-\left\vert S_{jj^{\prime}}\right\vert ^{2}e^{i\omega
^{+}\left(  \frac{x}{v_{Fj}}+\frac{x^{\prime}}{v_{Fj^{\prime}}}\right)
}\right]  \text{.}%
\end{align}
We note that the $\left\vert S_{j^{\prime}j}\right\vert ^{2}$ term vanishes
because the associated singularities\ are on the wrong side of the contour.
Now combine the $n_{F}\left(  \epsilon_{2}\right)  $ and $n_{F}\left(
\epsilon_{1}\right)  $ terms and restore the cutoff $D$, recalling that
$\epsilon_{2}-\epsilon_{1}=\omega^{+}$. This gives%

\begin{equation}
\Omega_{jj^{\prime}}^{\left(  0\right)  }\left(  x,x^{\prime};\omega
^{+}\right)  =i\frac{e^{2}}{2\pi}\int d\epsilon_{2}\left[  n_{F}\left(
\epsilon_{2}\right)  -n_{F}\left(  \epsilon_{2}-\omega^{+}\right)  \right]
\left[  \delta_{jj^{\prime}}e^{i\omega^{+}\left\vert \frac{x}{v_{Fj}}%
-\frac{x^{\prime}}{v_{Fj^{\prime}}}\right\vert }-0-\left\vert S_{jj^{\prime}%
}\right\vert ^{2}e^{i\omega^{+}\left(  \frac{x}{v_{Fj}}+\frac{x^{\prime}%
}{v_{Fj^{\prime}}}\right)  }\right]  \text{.} \label{zerothOmega3}%
\end{equation}
Substituting into Eq.~(\ref{Kubo}), taking the $\eta_{\omega}\rightarrow0^{+}$
limit and then the $\omega\rightarrow0$ limit, we obtain Eq.~(\ref{bubble}).

\subsection{First order\label{subsec:conductancefirst}}

At the first order, as shown in Fig.~\ref{fig:Oalpha}, the bubble diagram is
dressed by two types of self-energies: contraction of $E_{1}$ with $E_{2}$ or
$E_{3}$ with $E_{4}$ in Eq.~(\ref{interaction}) (the ``tadpole''), and
contraction of $E_{1}$ with $E_{4}$ or $E_{2}$ with $E_{3}$. In addition,
there are two types of first order vertex correction diagrams, the ``cracked
egg'' diagram and the ring diagram.

For the self-energy diagrams and the dressed conductance bubbles we need two
more types of Matsubara frequency sums. The first one is%

\begin{equation}
\frac{1}{\beta}\sum_{i\omega_{n}}\mathcal{G}_{j}\left(  \epsilon,i\omega
_{n}\right)  =n_{F}\left(  \epsilon\right)  \text{.} \label{notadpole}%
\end{equation}
The second one is%

\begin{align}
&  \frac{1}{\beta}\sum_{i\omega_{n}}\mathcal{G}_{j_{1}^{\prime}}\left(
\epsilon_{1}^{\prime},i\omega_{n}\right)  \mathcal{G}_{j_{1}}\left(
\epsilon_{1},i\omega_{n}\right)  \mathcal{G}_{j_{2}}\left(  \epsilon
_{2},i\omega_{n}+ip_{m}\right) \nonumber\\
&  =n_{F}\left(  \epsilon_{2}\right)  \frac{1}{\left(  \epsilon_{2}%
-ip_{m}\right)  -\epsilon_{1}^{\prime}}\frac{1}{\left(  \epsilon_{2}%
-ip_{m}\right)  -\epsilon_{1}}-\int\frac{d\tilde{\epsilon}}{2\pi i}%
n_{F}\left(  \tilde{\epsilon}\right) \nonumber\\
&  \times\frac{1}{\tilde{\epsilon}+ip_{m}-\epsilon_{2}}\left[  \frac{1}%
{\tilde{\epsilon}+i0-\epsilon_{1}^{\prime}}\frac{1}{\tilde{\epsilon
}+i0-\epsilon_{1}}-\frac{1}{\tilde{\epsilon}-i0-\epsilon_{1}^{\prime}}\frac
{1}{\tilde{\epsilon}-i0-\epsilon_{1}}\right]  \text{.} \label{branchcut1}%
\end{align}
To compute this sum, we consider the following contour integral,%

\begin{equation}
\oint\frac{dz}{2\pi i}n_{F}\left(  z\right)  \frac{1}{z-\epsilon_{1}^{\prime}%
}\frac{1}{z-\epsilon_{1}}\frac{1}{z+ip_{m}-\epsilon_{2}}\text{,}%
\end{equation}
where the integration contour is wrapped around the branch cut on the real
axis,\cite{mahan2000many} so that poles inside the contour are $z=i\omega_{n}$
($n$ running over all integers)\ and also $z=\epsilon_{2}-ip_{m}$. The
$n_{F}\left(  \epsilon_{2}\right)  $ term in Eq.~(\ref{branchcut1}) comes from
$z=\epsilon_{2}-ip_{m}$, and the $n_{F}\left(  \tilde{\epsilon}\right)  $ term
comes from the branch cut $z=0$.

We ignore the tadpole-type self-energy diagrams, again on the grounds that
they only modify the chemical potential. The other type of self-energy
diagrams turn out to dress the S-matrix as in Eq.~(\ref{OalphaSdress}). One
instance of these diagrams reads%

\begin{align}
&  \Omega_{jj^{\prime}}^{\left(  1\right)  \text{,SE,non-tadpole,1}}\left(
x,x^{\prime};\tau-\tau^{\prime}\right)  =-\frac{e^{2}}{\left(  2\pi\right)
^{2}}\sum_{j_{1}j_{2}j_{1}^{\prime}j_{2}^{\prime}}\int d\epsilon_{1}%
d\epsilon_{2}d\epsilon_{1}^{\prime}d\epsilon_{2}^{\prime}\nonumber\\
&  \times\left[  e^{i\frac{\left(  \epsilon_{2}-\epsilon_{1}\right)  }{v_{Fj}%
}x}S_{jj_{1}}^{\ast}S_{jj_{2}}-e^{-i\frac{\left(  \epsilon_{2}-\epsilon
_{1}\right)  }{v_{Fj}}x}\delta_{jj_{1}}\delta_{jj_{2}}\right]  \left[
e^{i\frac{\left(  \epsilon_{2}^{\prime}-\epsilon_{1}^{\prime}\right)
}{v_{Fj^{\prime}}}x^{\prime}}S_{j^{\prime}j_{1}^{\prime}}^{\ast}S_{j^{\prime
}j_{2}^{\prime}}-e^{-i\frac{\left(  \epsilon_{2}^{\prime}-\epsilon_{1}%
^{\prime}\right)  }{v_{Fj^{\prime}}}x^{\prime}}\delta_{j^{\prime}j_{1}%
^{\prime}}\delta_{j^{\prime}j_{2}^{\prime}}\right] \nonumber\\
&  \times\left(  -\right)  \int_{0}^{\beta}d\tau_{1}\sum_{n}\int dy\,g_{2}%
^{n}\left(  y\right)  \sum_{l_{1}l_{2}l_{3}l_{4}}\int\frac{dE_{1}dE_{2}%
dE_{3}dE_{4}}{\left(  2\pi\right)  ^{2}v_{Fn}^{2}}\varrho_{l_{1}l_{2}%
l_{3}l_{4}}^{n}\left(  E_{1},E_{2},E_{3},E_{4};y\right) \nonumber\\
&  \times\delta_{j_{2}j_{1}^{\prime}}\delta\left(  \epsilon_{2}-\epsilon
_{1}^{\prime}\right)  \mathcal{G}_{j_{2}}\left(  \epsilon_{2},\tau
-\tau^{\prime}\right)  \delta_{j_{2}^{\prime}l_{1}}\delta\left(  \epsilon
_{2}^{\prime}-E_{1}\right)  \mathcal{G}_{j_{2}^{\prime}}\left(  \epsilon
_{2}^{\prime},\tau^{\prime}-\tau_{1}\right) \nonumber\\
&  \times\delta_{l_{2}l_{3}}\delta\left(  E_{2}-E_{3}\right)  \mathcal{G}%
_{l_{2}}\left(  E_{2},0\right)  \delta_{l_{4}j_{1}}\delta\left(
E_{4}-\epsilon_{1}\right)  \mathcal{G}_{j_{1}}\left(  \epsilon_{1},\tau
_{1}-\tau\right)  \text{.}%
\end{align}
Going to the frequency space, performing Matsubara sums and analytic
continuation, we find%

\begin{align}
&  \Omega_{jj^{\prime}}^{\left(  1\right)  \text{,SE,non-tadpole,1}}\left(
x,x^{\prime};\omega^{+}\right)  =-\frac{e^{2}}{\left(  2\pi\right)  ^{2}}%
\sum_{j_{1}j_{2}j_{2}^{\prime}}\int d\epsilon_{1}d\epsilon_{2}d\epsilon
_{2}^{\prime}\nonumber\\
&  \times\left[  e^{i\frac{\left(  \epsilon_{2}-\epsilon_{1}\right)  }{v_{Fj}%
}x}S_{jj_{1}}^{\ast}S_{jj_{2}}-e^{-i\frac{\left(  \epsilon_{2}-\epsilon
_{1}\right)  }{v_{Fj}}x}\delta_{jj_{1}}\delta_{jj_{2}}\right]  \left[
e^{i\frac{\left(  \epsilon_{2}^{\prime}-\epsilon_{2}\right)  }{v_{Fj^{\prime}%
}}x^{\prime}}S_{j^{\prime}j_{2}}^{\ast}S_{j^{\prime}j_{2}^{\prime}}%
-e^{-i\frac{\left(  \epsilon_{2}^{\prime}-\epsilon_{2}\right)  }%
{v_{Fj^{\prime}}}x^{\prime}}\delta_{j^{\prime}j_{2}}\delta_{j^{\prime}%
j_{2}^{\prime}}\right] \nonumber\\
&  \times\left(  -\right)  \sum_{n}\int dy\,g_{2}^{n}\left(  y\right)
\int\frac{dE_{2}}{\left(  2\pi\right)  ^{2}v_{Fn}^{2}}n_{F}\left(
E_{2}\right)  \varrho_{j_{2}^{\prime}nnj_{1}}^{n}\left(  \epsilon_{2}^{\prime
},E_{2},E_{2},\epsilon_{1};y\right) \nonumber\\
&  \times\left[  n_{F}\left(  \epsilon_{2}\right)  \frac{1}{\left(
\epsilon_{2}-\omega^{+}\right)  -\epsilon_{2}^{\prime}}\frac{1}{\left(
\epsilon_{2}-\omega^{+}\right)  -\epsilon_{1}}-\int\frac{d\tilde{\epsilon}%
}{2\pi i}n_{F}\left(  \tilde{\epsilon}\right)  \frac{1}{\tilde{\epsilon
}+\omega^{+}-\epsilon_{2}}\right. \nonumber\\
&  \left.  \times\left(  \frac{1}{\tilde{\epsilon}+i0-\epsilon_{2}^{\prime}%
}\frac{1}{\tilde{\epsilon}+i0-\epsilon_{1}}-\frac{1}{\tilde{\epsilon
}-i0-\epsilon_{2}^{\prime}}\frac{1}{\tilde{\epsilon}-i0-\epsilon_{1}}\right)
\right]  \text{.}%
\end{align}
Carrying out the $\epsilon_{1}$ and $\epsilon_{2}^{\prime}$ integrations, and
also the $\epsilon_{2}$ integration in the $n_{F}\left(  \tilde{\epsilon
}\right)  $ term, this becomes in the $x$, $x^{\prime}\rightarrow\infty$ limit%

\begin{align}
&  \Omega_{jj^{\prime}}^{\left(  1\right)  \text{,SE,non-tadpole,1}}\left(
x,x^{\prime};\omega^{+}\right)  =-\frac{e^{2}}{2\pi}\int d\epsilon_{2}\left[
n_{F}\left(  \epsilon_{2}\right)  -n_{F}\left(  \epsilon_{2}-\omega
^{+}\right)  \right] \nonumber\\
&  \times\left(  -\right)  \sum_{n}\int dy\,\alpha_{n}\left(  y\right)
\int\frac{dE_{2}}{v_{Fn}}n_{F}\left(  E_{2}\right)  \frac{1}{2}e^{i\omega
^{+}\left(  \frac{x}{v_{Fj}}+\frac{x^{\prime}}{v_{Fj^{\prime}}}\right)
}\nonumber\\
&  \times S_{jj^{\prime}}\left[  S_{jn}^{\ast}S_{nn}S_{nj^{\prime}}^{\ast
}e^{2i\left(  E_{2}-\left(  \epsilon_{2}-\omega^{+}\right)  \right)  \frac
{y}{v_{Fn}}}+\delta_{jn}S_{nn}^{\ast}\delta_{j^{\prime}n}e^{-2i\left(
E_{2}-\left(  \epsilon_{2}-\omega^{+}\right)  \right)  \frac{y}{v_{Fn}}%
}\right]  \text{.}%
\end{align}
This is just one of the four terms which reproduce Eqs.~(\ref{OalphaFL}) and
(\ref{OalphaSdress0}) when inserted in Eq.~(\ref{Kubo}). Another identical
term comes from contracting $E_{1}$ with $E_{4}$ (completely equivalent to
contracting $E_{2}$ with $E_{3}$ which we have done). The remaining two terms
have all their electron propagators reverted, so that their contributions to
the conductance are the complex conjugate of the first two terms. This
concludes the derivation of Eqs.~(\ref{OalphaFL}) and (\ref{OalphaSdress0}).

Neither type of vertex corrections to the conductance requires Matsubara sums
other than Eq.~(\ref{bubbleMats}). An example of the ring diagram is%

\begin{align}
&  \Omega_{jj^{\prime}}^{\left(  1\right)  \text{,VC,ring,1}}\left(
x,x^{\prime};\tau-\tau^{\prime}\right)  =-\frac{e^{2}}{\left(  2\pi\right)
^{2}}\sum_{j_{1}j_{2}j_{1}^{\prime}j_{2}^{\prime}}\int d\epsilon_{1}%
d\epsilon_{2}d\epsilon_{1}^{\prime}d\epsilon_{2}^{\prime}\nonumber\\
&  \times\left[  e^{i\frac{\left(  \epsilon_{2}-\epsilon_{1}\right)  }{v_{Fj}%
}x}S_{jj_{1}}^{\ast}S_{jj_{2}}-e^{-i\frac{\left(  \epsilon_{2}-\epsilon
_{1}\right)  }{v_{Fj}}x}\delta_{jj_{1}}\delta_{jj_{2}}\right]  \left[
e^{i\frac{\left(  \epsilon_{2}^{\prime}-\epsilon_{1}^{\prime}\right)
}{v_{Fj^{\prime}}}x^{\prime}}S_{j^{\prime}j_{1}^{\prime}}^{\ast}S_{j^{\prime
}j_{2}^{\prime}}-e^{-i\frac{\left(  \epsilon_{2}^{\prime}-\epsilon_{1}%
^{\prime}\right)  }{v_{Fj^{\prime}}}x^{\prime}}\delta_{j^{\prime}j_{1}%
^{\prime}}\delta_{j^{\prime}j_{2}^{\prime}}\right] \nonumber\\
&  \times\left(  -\right)  \int_{0}^{\beta}d\tau_{1}\sum_{n}\int dy\,g_{2}%
^{n}\left(  y\right)  \sum_{l_{1}l_{2}l_{3}l_{4}}\int\frac{dE_{1}dE_{2}%
dE_{3}dE_{4}}{\left(  2\pi\right)  ^{2}v_{Fn}^{2}}\varrho_{l_{1}l_{2}%
l_{3}l_{4}}^{n}\left(  E_{1},E_{2},E_{3},E_{4};y\right) \nonumber\\
&  \times\delta_{j_{2}l_{1}}\delta\left(  \epsilon_{2}-E_{1}\right)
\mathcal{G}_{j_{2}}\left(  \epsilon_{2},\tau-\tau_{1}\right)  \delta
_{l_{4}j_{1}^{\prime}}\delta\left(  \epsilon_{1}^{\prime}-E_{4}\right)
\mathcal{G}_{j_{1}^{\prime}}\left(  \epsilon_{1}^{\prime},\tau_{1}%
-\tau^{\prime}\right) \nonumber\\
&  \times\delta_{j_{2}^{\prime}l_{3}}\delta\left(  \epsilon_{2}^{\prime}%
-E_{3}\right)  \mathcal{G}_{j_{2}^{\prime}}\left(  \epsilon_{2}^{\prime}%
,\tau^{\prime}-\tau_{1}\right)  \delta_{j_{1}l_{2}}\delta\left(
E_{2}-\epsilon_{1}\right)  \mathcal{G}_{j_{1}}\left(  \epsilon_{1},\tau
_{1}-\tau\right)  \text{.}%
\end{align}
Going to the frequency space, performing Matsubara sums and analytic continuation:%

\begin{align}
&  \Omega_{jj^{\prime}}^{\left(  1\right)  \text{,VC,ring,1}}\left(
x,x^{\prime};\omega^{+}\right)  =-\frac{e^{2}}{\left(  2\pi\right)  ^{2}}%
\sum_{j_{1}j_{2}j_{1}^{\prime}j_{2}^{\prime}}\int d\epsilon_{1}d\epsilon
_{2}d\epsilon_{1}^{\prime}d\epsilon_{2}^{\prime}\nonumber\\
&  \times\left[  e^{i\frac{\left(  \epsilon_{2}-\epsilon_{1}\right)  }{v_{Fj}%
}x}S_{jj_{1}}^{\ast}S_{jj_{2}}-e^{-i\frac{\left(  \epsilon_{2}-\epsilon
_{1}\right)  }{v_{Fj}}x}\delta_{jj_{1}}\delta_{jj_{2}}\right]  \left[
e^{i\frac{\left(  \epsilon_{2}^{\prime}-\epsilon_{1}^{\prime}\right)
}{v_{Fj^{\prime}}}x^{\prime}}S_{j^{\prime}j_{1}^{\prime}}^{\ast}S_{j^{\prime
}j_{2}^{\prime}}-e^{-i\frac{\left(  \epsilon_{2}^{\prime}-\epsilon_{1}%
^{\prime}\right)  }{v_{Fj^{\prime}}}x^{\prime}}\delta_{j^{\prime}j_{1}%
^{\prime}}\delta_{j^{\prime}j_{2}^{\prime}}\right] \nonumber\\
&  \times\sum_{n}\int dy\,g_{2}^{n}\left(  y\right)  \frac{1}{\left(
2\pi\right)  ^{2}v_{Fn}^{2}}\varrho_{j_{2}j_{1}j_{2}^{\prime}j_{1}^{\prime}%
}^{n}\left(  \epsilon_{2},\epsilon_{1},\epsilon_{2}^{\prime},\epsilon
_{1}^{\prime};y\right)  \frac{n_{F}\left(  \epsilon_{2}\right)  -n_{F}\left(
\epsilon_{1}\right)  }{\epsilon_{2}-\epsilon_{1}-\omega^{+}}\frac{n_{F}\left(
\epsilon_{2}^{\prime}\right)  -n_{F}\left(  \epsilon_{1}^{\prime}\right)
}{\epsilon_{2}^{\prime}-\epsilon_{1}^{\prime}+\omega^{+}}\text{.}%
\end{align}
Integrating over the energies as before, we find%

\begin{align}
&  \Omega_{jj^{\prime}}^{\left(  1\right)  \text{,VC,ring,1}}\left(
x,x^{\prime};\omega^{+}\right)  =-\frac{e^{2}}{2\pi}\int d\epsilon
_{1}d\epsilon_{1}^{\prime}\left(  -\right)  \left[  n_{F}\left(  \epsilon
_{1}+\omega^{+}\right)  -n_{F}\left(  \epsilon_{1}\right)  \right]  \left[
n_{F}\left(  \epsilon_{1}^{\prime}-\omega^{+}\right)  -n_{F}\left(
\epsilon_{1}^{\prime}\right)  \right] \nonumber\\
&  \times\sum_{n}\int dy\,\alpha_{n}\left(  y_{1}\right)  \frac{1}{v_{Fn}%
}\frac{1}{2}\left[  \left\vert S_{jn}\right\vert ^{2}\delta_{j^{\prime}%
n}e^{i\omega^{+}\left(  \frac{x}{v_{Fj}}-\frac{x^{\prime}}{v_{Fj^{\prime}}%
}\right)  }e^{2i\omega^{+}\frac{y}{v_{Fn}}}\theta\left(  \frac{y}{v_{Fn}%
}-\frac{x^{\prime}}{v_{Fj^{\prime}}}\right)  \right. \nonumber\\
&  \left.  -\delta_{jn}\delta_{j^{\prime}n}e^{-i\omega^{+}\left(  \frac
{x}{v_{Fj}}+\frac{x^{\prime}}{v_{Fj^{\prime}}}\right)  }e^{2i\omega^{+}%
\frac{y}{v_{Fn}}}\theta\left(  \frac{y}{v_{Fn}}-\frac{x}{v_{Fj}}\right)
\theta\left(  \frac{y}{v_{Fn}}-\frac{x^{\prime}}{v_{Fj^{\prime}}}\right)
\right. \nonumber\\
&  \left.  -\delta_{jn}\delta_{j^{\prime}n}e^{i\omega^{+}\left(  \frac
{x}{v_{Fj}}+\frac{x^{\prime}}{v_{Fj^{\prime}}}\right)  }e^{-2i\omega^{+}%
\frac{y}{v_{Fn}}}\theta\left(  \frac{x}{v_{Fj}}-\frac{y}{v_{Fn}}\right)
\theta\left(  \frac{x^{\prime}}{v_{Fj^{\prime}}}-\frac{y}{v_{Fn}}\right)
\right. \nonumber\\
&  \left.  -\left\vert S_{jn}\right\vert ^{2}\left\vert S_{nj^{\prime}%
}\right\vert ^{2}e^{i\omega^{+}\left(  \frac{x}{v_{Fj}}+\frac{x^{\prime}%
}{v_{Fj^{\prime}}}\right)  }e^{2i\omega^{+}\frac{y}{v_{Fn}}}+\delta
_{jn}\left\vert S_{nj^{\prime}}\right\vert ^{2}e^{-i\omega^{+}\left(  \frac
{x}{v_{Fj}}-\frac{x^{\prime}}{v_{Fj^{\prime}}}\right)  }e^{2i\omega^{+}%
\frac{y}{v_{Fn}}}\theta\left(  \frac{y}{v_{Fn}}-\frac{x}{v_{Fj}}\right)
\right]  \text{.}%
\end{align}
There exists an analogous term with all electron lines reverted. Upon
substitution into Eq.~(\ref{Kubo}) these two terms reproduce Eq.~(\ref{OalphaTLL0}).

Finally, by summing over all dummy wire indices, we can show that the ``cracked
egg'' contribution to the DC conductance is proportional to $\delta
_{jj^{\prime}}$. On the other hand, due to current conservation and the
absence of equilibrium currents, the full DC conductance $G_{jj^{\prime}}$ obeys%

\begin{equation}
\sum_{j}G_{jj^{\prime}}=\sum_{j^{\prime}}G_{jj^{\prime}}=0\text{;}
\label{Growsum}%
\end{equation}
this must also be true at $O\left(  \alpha\right)  $. Since Eq.~(\ref{Growsum}) is already satisfied by Eqs.~(\ref{OalphaFL}) and (\ref{OalphaTLL}), it must also be separately satisfied by the ``cracked egg''
diagrams. But $\sum_{j}\delta_{jj^{\prime}}=\sum_{j^{\prime}}\delta
_{jj^{\prime}}=1$, and we infer that the ``cracked egg'' diagrams must be
identically zero.

\section{Details of the Wilsonian approach to S-matrix renormalization\label{sec:appOalphaNG}}

In this appendix, we review the derivation of the S-matrix RG equation using
the Wilsonian scaling approach in Ref.~\onlinecite{PhysRevB.49.1966}.

Starting from Eq.~(\ref{interaction}), we reduce the energy cutoff $D$ to
$D-\delta D$ ($\delta D\ll D$), and integrate out the so-called ``fast modes''
with energies in one of the two slices $\left(  -D+\delta D,-D\right)  $ and
$\left(  D-\delta D,D\right)  $. This procedure generates corrections of
$O\left(  \alpha\delta D/D\right)  $ to the quadratic part of the Hamiltonian
[Eq.~(\ref{quadratic})] as well as the quartic part [Eq.~(\ref{interaction})]. We assume that the corrections to the quartic part are
unimportant; the rationale is that the quartic part originates entirely from
the bulk, so it should renormalize independently of the junction. In fact,
since the quartic part is free of Umklapp processes, it should be exactly
marginal in the RG sense.\cite{RevModPhys.66.129} Meanwhile, the renormalized
quadratic part becomes off-diagonal and must be diagonalized with a new
scattering basis, which is in turn associated with a running (i.e.
cutoff-dependent) S-matrix.

The quadratic correction generated by Eq.~(\ref{interaction}) reads%

\begin{equation}
\delta H_{0}^{j}=-2\int_{0}^{\infty}dxg_{2}^{j}\left(  x\right)  \sum
_{l_{1}l_{2}l_{3}}\int_{\delta D}dE_{1}\int_{-D}^{D}\frac{dE_{2}dE_{3}%
}{\left(  2\pi\right)  ^{2}v_{Fj}^{2}}\varrho_{l_{1}l_{2}l_{3}l_{1}}%
^{j}\left(  E_{1},E_{2},E_{3},E_{1};x\right)  n_{F}\left(  E_{1}\right)
\phi_{l_{3}}^{\dagger}\left(  E_{3}\right)  \phi_{l_{2}}\left(  E_{2}\right)
\text{.}%
\end{equation}
The $E_{2}E_{3}$ contraction is equivalent to the $E_{1}E_{4}$ contraction;
hence the factor of $2$. The $E_{1}E_{2}$ and $E_{3}E_{4}$ contractions are
discarded because, once we sum over $l$ taking into account the S-matrix
unitarity $\sum_{l_{1}}\left\vert S_{jl_{1}}\right\vert ^{2}=1$, we find they
only harmlessly shift the chemical potential.\cite{RevModPhys.66.129}

We let $\phi^{\prime}$ be the renormalized scattering basis after integrating
out fast modes. $\phi^{\prime}$ is related to $\phi$ by another S-matrix,
$S_{jj^{\prime}}^{\delta}$,\ which only weakly deviates from the $N\times N$
identity matrix:%

\begin{equation}
\phi_{j}\left(  E\right)  =\int\frac{dE^{\prime}}{2\pi}\left[  \frac
{i}{E-E^{\prime}+i0}\phi_{j}^{\prime}\left(  E^{\prime}\right)  +\frac
{-i}{E-E^{\prime}-i0}\sum_{j^{\prime}}S_{jj^{\prime}}^{\delta}\left(
E^{\prime};E\right)  \phi_{j^{\prime}}^{\prime}\left(  E^{\prime}\right)
\right]  \label{newscatbas1}%
\end{equation}
The inverse transformation is obtained by calculating anti-commutators:%

\begin{equation}
\left(  \phi_{j^{\prime}}^{\prime}\left(  E^{\prime}\right)  \right)  ^{\dag
}=\sum_{j}\int\frac{dE}{2\pi}\left[  \frac{i}{E-E^{\prime}+i0}\delta
_{jj^{\prime}}+\frac{-i}{E-E^{\prime}-i0}S_{jj^{\prime}}^{\delta}\left(
E^{\prime};E\right)  \right]  \phi_{j}^{\dag}\left(  E\right)
\label{newscatbas2}%
\end{equation}

By definition $\phi^{\prime}$ diagonalizes the renormalized quadratic Hamiltonian,%

\begin{equation}
\left[  \sum_{j}H_{0\text{,wire}}^{j}+H_{0,B}+\sum_{j}\delta H_{0}^{j},\left(
\phi_{j^{\prime}}^{\prime}\left(  E^{\prime}\right)  \right)  ^{\dag}\right]
=E^{\prime}\left(  \phi_{j^{\prime}}^{\prime}\left(  E^{\prime}\right)
\right)  ^{\dag}%
\end{equation}
Substituting Eq.~(\ref{newscatbas2}) into the above, we find to $O\left(
\delta D/D\right)  $%

\begin{align}
&  i\left[  \delta_{jj^{\prime}}-S_{jj^{\prime}}^{\delta}\left(  E^{\prime
};E\right)  \right]  =\delta D\int_{0}^{\infty}dx\left\{  n_{F}\left(
D\right)  \left[  \frac{\alpha_{j}\left(  x\right)  }{v_{Fj}}S_{jj}^{\ast
}S_{jj^{\prime}}e^{-i\frac{2D-E-E^{\prime}}{v_{Fj}}x}+\frac{\alpha_{j^{\prime
}}\left(  x\right)  }{v_{Fj^{\prime}}}S_{j^{\prime}j^{\prime}}S_{j^{\prime}%
j}^{\ast}e^{i\frac{2D-E-E^{\prime}}{v_{Fj^{\prime}}}x}\right]  \right.
\nonumber\\
&  \left.  +n_{F}\left(  -D\right)  \left[  \frac{\alpha_{j}\left(  x\right)
}{v_{Fj}}S_{jj}^{\ast}S_{jj^{\prime}}e^{i\frac{2D+E+E^{\prime}}{v_{Fj}}%
x}+\frac{\alpha_{j^{\prime}}\left(  x\right)  }{v_{Fj^{\prime}}}S_{j^{\prime
}j^{\prime}}S_{j^{\prime}j}^{\ast}e^{-i\frac{2D+E+E^{\prime}}{v_{Fj^{\prime}}%
}x}\right]  \right\}
\end{align}
For the simple model Eq.~(\ref{interactionmodel}), integrating over $x$, we find%

\begin{align}
&  \delta_{jj^{\prime}}-S_{jj^{\prime}}^{\delta}\left(  E^{\prime};E\right)
=\delta D\left\{  n_{F}\left(  D\right)  \left[  S_{jj}^{\ast}S_{jj^{\prime}%
}\frac{\left(  \alpha_{j}\left(  0\right)  -\alpha_{j}\left(  \infty\right)
\right)  e^{-i\frac{2D-E-E^{\prime}}{v_{Fj}}L_{j}}-\alpha_{j}\left(  0\right)
}{2D-E-E^{\prime}}\right.  \right. \nonumber\\
&  \left.  \left.  +S_{j^{\prime}j^{\prime}}S_{j^{\prime}j}^{\ast}%
\frac{\left(  \alpha_{j^{\prime}}\left(  0\right)  -\alpha_{j^{\prime}}\left(
\infty\right)  \right)  e^{i\frac{2D-E-E^{\prime}}{v_{Fj^{\prime}}%
}L_{j^{\prime}}}-\alpha_{j^{\prime}}\left(  0\right)  }{2D-E-E^{\prime}%
}\right]  +n_{F}\left(  -D\right)  \left[  S_{jj}^{\ast}S_{jj^{\prime}}%
\frac{\left(  \alpha_{j}\left(  0\right)  -\alpha_{j}\left(  \infty\right)
\right)  e^{i\frac{2D+E+E^{\prime}}{v_{Fj}}L_{j}}-\alpha_{j}\left(  0\right)
}{-2D-E-E^{\prime}}\right.  \right. \nonumber\\
&  \left.  \left.  +S_{j^{\prime}j^{\prime}}S_{j^{\prime}j}^{\ast}%
\frac{\left(  \alpha_{j^{\prime}}\left(  0\right)  -\alpha_{j^{\prime}}\left(
\infty\right)  \right)  e^{-i\frac{2D+E+E^{\prime}}{v_{Fj^{\prime}}%
}L_{j^{\prime}}}-\alpha_{j^{\prime}}\left(  0\right)  }{-2D-E-E^{\prime}%
}\right]  \right\}  \label{Sdelta0}%
\end{align}
When we assume $D\gtrsim\max\left\{  \left\vert E\right\vert ,\left\vert
E^{\prime}\right\vert ,T\right\}  $, and apply the same considerations below
Eq.~(\ref{Srenor0}), Eq.~(\ref{Sdelta0}) becomes%

\begin{equation}
S_{jj^{\prime}}^{\delta}\left(  E;E^{\prime}\right)  =\delta_{jj^{\prime}%
}-\frac{\delta D}{2D}\left(  \alpha_{j}\left(  D\right)  S_{jj}^{\ast
}S_{jj^{\prime}}-\alpha_{j^{\prime}}\left(  D\right)  S_{j^{\prime}j^{\prime}%
}S_{j^{\prime}j}^{\ast}\right)  \text{.} \label{Sdelta}%
\end{equation}

The renormalized S-matrix $S+\delta S$ relates $\phi^{\prime}$ to the original
fermions $\psi$. Inserting Eq.~(\ref{newscatbas1}) into Eq.~(\ref{scatbas}) we
find that $\delta S$ and $S^{\delta}$ obey the simple matrix relation $\delta
S=SS^{\delta}-S$, and according to Eq.~(\ref{Sdelta}), $\delta S_{jj^{\prime}%
}$ is given by none other than Eq.~(\ref{Srenor}). Thus to the first order in
interaction the CS approach and the Wilsonian approach predict the same
S-matrix renormalization, Eq.~(\ref{NazarovGlazman}).

\section{Details of the RPA\label{sec:appRPA}}

In this appendix we expound the RPA calculations that lead to Eqs.~(\ref{RPANazarovGlazman}) and (\ref{RPATLL}).

\subsection{RPA conductance\label{subsec:conductanceRPA}}

The RPA self-energy beyond the first order involves a new type of Matsubara
sum. For instance, at the third order in interaction, we need%

\begin{align}
&  \frac{1}{\beta}\sum_{ip_{m}}\frac{n_{F}\left(  E_{4}\right)  -n_{F}\left(
E_{3}\right)  }{ip_{m}+E_{4}-E_{3}}\frac{n_{F}\left(  E_{8}\right)
-n_{F}\left(  E_{7}\right)  }{ip_{m}+E_{8}-E_{7}}\frac{1}{i\left(
p_{m}+\omega_{n}\right)  -E_{2}}\nonumber\\
&  =n_{F}\left(  E_{2}\right)  \frac{n_{F}\left(  E_{4}\right)  -n_{F}\left(
E_{3}\right)  }{E_{2}-i\omega_{n}+E_{4}-E_{3}}\frac{n_{F}\left(  E_{8}\right)
-n_{F}\left(  E_{7}\right)  }{E_{2}-i\omega_{n}+E_{8}-E_{7}}\nonumber\\
&  +\int\frac{d\tilde{\epsilon}}{2\pi i}n_{B}\left(  \tilde{\epsilon}\right)
\frac{1}{\tilde{\epsilon}+i\omega_{n}-E_{2}}\left(  \frac{n_{F}\left(
E_{4}\right)  -n_{F}\left(  E_{3}\right)  }{\left(  \tilde{\epsilon
}+i0\right)  +E_{4}-E_{3}}\frac{n_{F}\left(  E_{8}\right)  -n_{F}\left(
E_{7}\right)  }{\left(  \tilde{\epsilon}+i0\right)  +E_{8}-E_{7}}\right.
\nonumber\\
&  \left.  -\frac{n_{F}\left(  E_{4}\right)  -n_{F}\left(  E_{3}\right)
}{\left(  \tilde{\epsilon}-i0\right)  +E_{4}-E_{3}}\frac{n_{F}\left(
E_{8}\right)  -n_{F}\left(  E_{7}\right)  }{\left(  \tilde{\epsilon
}-i0\right)  +E_{8}-E_{7}}\right)  \label{branchcut2}%
\end{align}
where $n_{B}\left(  \epsilon\right)  =1/\left(  e^{\beta\epsilon}-1\right)  $
is the Bose distribution. To derive Eq.~(\ref{branchcut2}) we again wrap the
integration contour around the branch cut at the real axis. The fraction with
numerator $n_{F}\left(  E_{3}\right)  -n_{F}\left(  E_{4}\right)
$\ originates from the fermion loop with loop energy $E_{3}$ and $E_{4}$; at
the $l$th order there will be $l-1$ loops present. $ip_{m}$ is the bosonic
frequency carried by the interaction lines; after $ip_{m}$, $i\omega_{n}$ is
also summed over following Eq.~(\ref{branchcut1}).

After we perform analytic continuation and integrate over the loop momenta, as
$x$, $x^{\prime}\rightarrow\infty$, the three most important terms in the
correlation function at the third order are%

\begin{subequations}
\begin{align}
&  \Omega_{jj^{\prime}}^{\left(  3\right)  \text{,SE,RPA,}E_{2}}\left(
x,x^{\prime};\omega^{+}\right)  =\frac{e^{2}}{2\pi}e^{i\omega^{+}\left(
\frac{x}{v_{Fj}}+\frac{x^{\prime}}{v_{Fj^{\prime}}}\right)  }\int
d\epsilon_{1}^{\prime}\left[  n_{F}\left(  \epsilon_{1}^{\prime}\right)
-n_{F}\left(  \epsilon_{1}^{\prime}-\omega^{+}\right)  \right]  S_{jj^{\prime
}}\nonumber\\
&  \times\sum_{n_{1}n_{2}n_{3}}\int_{0}^{\infty}d\tilde{y}_{1}d\tilde{y}%
_{2}d\tilde{y}_{3}\alpha_{n_{1}}\left(  v_{Fn_{1}}\tilde{y}_{1}\right)
\alpha_{n_{2}}\left(  v_{Fn_{2}}\tilde{y}_{2}\right)  \alpha_{n_{3}}\left(
v_{Fn_{3}}\tilde{y}_{3}\right)  \int dE_{2}n_{F}\left(  E_{2}\right)
\tilde{E}_{2}^{2}\nonumber\\
&  \times\left[  S_{jn_{3}}^{\ast}S_{n_{1}n_{3}}S_{n_{1}j^{\prime}}^{\ast
}\left\vert S_{n_{2}n_{1}}\right\vert ^{2}\left\vert S_{n_{3}n_{2}}\right\vert
^{2}e^{2i\tilde{E}_{2}\left(  \tilde{y}_{1}+\tilde{y}_{2}+\tilde{y}%
_{3}\right)  }\right. \nonumber\\
&  \left.  +S_{jn_{3}}^{\ast}S_{n_{1}n_{3}}S_{n_{1}j^{\prime}}^{\ast}%
\delta_{n_{3}n_{2}}\delta_{n_{2}n_{1}}\theta\left(  \tilde{y}_{1}-\tilde
{y}_{2}\right)  \theta\left(  \tilde{y}_{3}-\tilde{y}_{2}\right)
e^{2i\tilde{E}_{2}\left(  \tilde{y}_{1}-\tilde{y}_{2}+\tilde{y}_{3}\right)
}\right. \nonumber\\
&  \left.  +\delta_{jn_{3}}\delta_{n_{3}n_{1}}S_{n_{1}j^{\prime}}^{\ast}%
\delta_{n_{3}n_{2}}\left\vert S_{n_{2}n_{1}}\right\vert ^{2}\theta\left(
\tilde{y}_{2}-\tilde{y}_{3}\right)  e^{2i\tilde{E}_{2}\left(  \tilde{y}%
_{1}+\tilde{y}_{2}-\tilde{y}_{3}\right)  }\right. \nonumber\\
&  \left.  +S_{jn_{3}}^{\ast}\delta_{n_{3}n_{1}}\delta_{n_{1}j^{\prime}%
}\left\vert S_{n_{3}n_{2}}\right\vert ^{2}\delta_{n_{2}n_{1}}\theta\left(
\tilde{y}_{2}-\tilde{y}_{1}\right)  e^{2i\tilde{E}_{2}\left(  -\tilde{y}%
_{1}+\tilde{y}_{2}+\tilde{y}_{3}\right)  }\right. \nonumber\\
&  \left.  +\delta_{jn_{3}}S_{n_{3}n_{1}}^{\ast}\delta_{n_{1}j^{\prime}}%
\delta_{n_{3}n_{2}}\delta_{n_{2}n_{1}}\theta\left(  \tilde{y}_{2}-\tilde
{y}_{1}\right)  \theta\left(  \tilde{y}_{2}-\tilde{y}_{3}\right)
e^{2i\tilde{E}_{2}\left(  -\tilde{y}_{1}+\tilde{y}_{2}-\tilde{y}_{3}\right)
}\right]  \label{thirdOmegaRPAq4}%
\end{align}
where we have substituted $\tilde{y}_{n}=y_{n}/v_{Fn}$ and $\tilde{E}%
_{2}=E_{2}-\epsilon_{1}^{\prime}+\omega^{+}$,%

\begin{align}
&  \Omega_{jj^{\prime}}^{\left(  3\right)  \text{,SE,RPA,}\tilde{\epsilon}%
+}\left(  x,x^{\prime};\omega^{+}\right)  =\frac{e^{2}}{2\pi}e^{i\omega
^{+}\left(  \frac{x}{v_{Fj}}+\frac{x^{\prime}}{v_{Fj^{\prime}}}\right)  }\int
d\epsilon_{1}^{\prime}\left[  n_{F}\left(  \epsilon_{1}^{\prime}\right)
-n_{F}\left(  \epsilon_{1}^{\prime}-\omega^{+}\right)  \right]  S_{jj^{\prime
}}\nonumber\\
&  \times\sum_{n_{1}n_{2}n_{3}}\int_{0}^{\infty}d\tilde{y}_{1}d\tilde{y}%
_{2}d\tilde{y}_{3}\alpha_{n_{1}}\left(  v_{Fn_{1}}\tilde{y}_{1}\right)
\alpha_{n_{2}}\left(  v_{Fn_{2}}\tilde{y}_{2}\right)  \alpha_{n_{3}}\left(
v_{Fn_{3}}\tilde{y}_{3}\right)  \int d\tilde{\epsilon}n_{B}\left(
\tilde{\epsilon}\right)  \tilde{\epsilon}^{2}\nonumber\\
&  \times\left[  \delta_{jn_{3}}\delta_{n_{3}n_{1}}S_{n_{1}j^{\prime}}^{\ast
}\delta_{n_{3}n_{2}}\left\vert S_{n_{2}n_{1}}\right\vert ^{2}\theta\left(
\tilde{y}_{2}-\tilde{y}_{3}\right)  \theta\left(  \tilde{y}_{3}-\tilde{y}%
_{1}\right)  e^{2i\left(  \tilde{\epsilon}+i0\right)  \left(  \tilde{y}%
_{1}+\tilde{y}_{2}-\tilde{y}_{3}\right)  }\right. \nonumber\\
&  \left.  +S_{jn_{3}}^{\ast}\delta_{n_{3}n_{1}}\delta_{n_{1}j^{\prime}%
}\left\vert S_{n_{3}n_{2}}\right\vert ^{2}\delta_{n_{2}n_{1}}\theta\left(
\tilde{y}_{2}-\tilde{y}_{1}\right)  \theta\left(  \tilde{y}_{1}-\tilde{y}%
_{3}\right)  e^{2i\left(  \tilde{\epsilon}+i0\right)  \left(  -\tilde{y}%
_{1}+\tilde{y}_{2}+\tilde{y}_{3}\right)  }\right. \nonumber\\
&  \left.  +\delta_{jn_{3}}S_{n_{3}n_{1}}^{\ast}\delta_{n_{1}j^{\prime}}%
\delta_{n_{3}n_{2}}\delta_{n_{2}n_{1}}\theta\left(  \tilde{y}_{2}-\tilde
{y}_{1}\right)  \theta\left(  \tilde{y}_{2}-\tilde{y}_{3}\right)  e^{2i\left(
\tilde{\epsilon}+i0\right)  \left(  -\tilde{y}_{1}+\tilde{y}_{2}-\tilde{y}%
_{3}\right)  }\right]  \label{thirdOmegaRPAe1p}%
\end{align}
and finally%

\begin{align}
&  \Omega_{jj^{\prime}}^{\left(  3\right)  \text{,SE,RPA,}\tilde{\epsilon}%
-}\left(  x,x^{\prime};\omega^{+}\right)  =\frac{e^{2}}{2\pi}e^{i\omega
^{+}\left(  \frac{x}{v_{Fj}}+\frac{x^{\prime}}{v_{Fj^{\prime}}}\right)  }\int
d\epsilon_{1}^{\prime}\left[  n_{F}\left(  \epsilon_{1}^{\prime}\right)
-n_{F}\left(  \epsilon_{1}^{\prime}-\omega^{+}\right)  \right]  S_{jj^{\prime
}}\nonumber\\
&  \times\sum_{n_{1}n_{2}n_{3}}\int_{0}^{\infty}d\tilde{y}_{1}d\tilde{y}%
_{2}d\tilde{y}_{3}\alpha_{n_{1}}\left(  v_{Fn_{1}}\tilde{y}_{1}\right)
\alpha_{n_{2}}\left(  v_{Fn_{2}}\tilde{y}_{2}\right)  \alpha_{n_{3}}\left(
v_{Fn_{3}}\tilde{y}_{3}\right)  \int d\tilde{\epsilon}n_{B}\left(
\tilde{\epsilon}\right)  \tilde{\epsilon}^{2}\nonumber\\
&  \times\left[  -\delta_{jn_{3}}\delta_{n_{3}n_{1}}S_{n_{1}j^{\prime}}^{\ast
}\delta_{n_{1}n_{2}}\left\vert S_{n_{2}n_{3}}\right\vert ^{2}\theta\left(
\tilde{y}_{2}-\tilde{y}_{1}\right)  \theta\left(  \tilde{y}_{3}-\tilde{y}%
_{1}\right)  e^{2i\left(  \tilde{\epsilon}-i0\right)  \left(  \tilde{y}%
_{1}-\tilde{y}_{2}-\tilde{y}_{3}\right)  }\right. \nonumber\\
&  \left.  -\delta_{jn_{3}}S_{n_{3}n_{1}}^{\ast}\delta_{n_{1}j^{\prime}}%
\delta_{n_{1}n_{2}}\delta_{n_{2}n_{3}}\theta\left(  \tilde{y}_{1}-\tilde
{y}_{2}\right)  \theta\left(  \tilde{y}_{3}-\tilde{y}_{2}\right)  e^{2i\left(
\tilde{\epsilon}-i0\right)  \left(  -\tilde{y}_{1}+\tilde{y}_{2}-\tilde{y}%
_{3}\right)  }\right. \nonumber\\
&  \left.  -S_{jn_{3}}^{\ast}\delta_{n_{3}n_{1}}\delta_{n_{1}j^{\prime}%
}\left\vert S_{n_{1}n_{2}}\right\vert ^{2}\delta_{n_{2}n_{3}}\theta\left(
\tilde{y}_{2}-\tilde{y}_{3}\right)  \theta\left(  \tilde{y}_{1}-\tilde{y}%
_{3}\right)  e^{2i\left(  \tilde{\epsilon}-i0\right)  \left(  -\tilde{y}%
_{1}-\tilde{y}_{2}+\tilde{y}_{3}\right)  }\right. \nonumber\\
&  \left.  -\delta_{jn_{3}}S_{n_{3}n_{1}}^{\ast}\delta_{n_{1}j^{\prime}%
}\left\vert S_{n_{1}n_{2}}\right\vert ^{2}\left\vert S_{n_{2}n_{3}}\right\vert
^{2}e^{2i\left(  \tilde{\epsilon}-i0\right)  \left(  -\tilde{y}_{1}-\tilde
{y}_{2}-\tilde{y}_{3}\right)  }\right]  \label{thirdOmegaRPAe1m}%
\end{align}
\end{subequations}
plus similar terms with all electron lines reverted. $\tilde{y}_{j}\equiv
y_{j}/v_{Fj}$ runs between $0$ and $\infty$, $j=1$, $2$, $3$. These three
terms come from lines 2, 3 and 4 of Eq.~(\ref{branchcut2}) respectively.

In the DC limit, the zeroth order contribution and the self-energy corrections
to the conductance again constitute a Landauer-type formula with a dressed
S-matrix, similar to Eq.~(\ref{OalphaFL}). Now we reduce the cutoff and demand
the conductance be cutoff-independent. Once the $\tilde{y}$ integrals are
performed, it is obvious that the cutoff-sensitive integrals are the $E_{2}$
integral and the $\tilde{\epsilon}$ integral.

We are in a position to discuss the real space integrals. We first focus on
the simplest case where the interactions in wires and leads are uniform and
identical, $\alpha_{n_{1}}\left(  y\right)  =$ $\alpha_{n_{1}}$ for any
$n_{1}$, so that all $\alpha$'s factor out. At the third order, we find the
following integrals:%

\begin{equation}
I_{1}\left(  E^{+}\right)  \equiv\int_{0}^{\infty}d\tilde{y}_{1}d\tilde{y}%
_{2}d\tilde{y}_{3}e^{iE^{+}\left(  \tilde{y}_{1}+\tilde{y}_{3}\right)
}e^{iE^{+}\left(  \tilde{y}_{1}-2\tilde{y}_{2}+\tilde{y}_{3}\right)  }%
\theta\left(  \tilde{y}_{1}-\tilde{y}_{2}\right)  \theta\left(  \tilde{y}%
_{3}-\tilde{y}_{2}\right)
\end{equation}
which appears alongside the factors $\delta_{n_{1}n_{2}}\delta_{n_{2}n_{3}}$, and%

\begin{equation}
\int_{0}^{\infty}d\tilde{y}_{1}e^{2iE^{+}\tilde{y}_{1}}=\frac{i}{2E^{+}}
\label{ordrsint}%
\end{equation}
which appears alongside, for example, $W_{n_{1}n_{2}}W_{n_{2}n_{3}}$. [More
accurately, Eq.~(\ref{ordrsint}) comes with each ``node'' $n_{2}$ as long as
$n_{2}$ is not sandwiched between two Kronecker $\delta$ factors.] Here
$E^{+}\equiv E+i0$ may be replaced by $\tilde{E}_{2}$ or\ $\left(  \pm
\tilde{\epsilon}+i0\right)  $. At higher orders, we need to evaluate the integral%

\begin{equation}
I_{M}\left(  E^{+}\right)  \equiv\prod_{l=1}^{2M+1}\left(  \int_{0}^{\infty
}d\tilde{y}_{l}\right)  e^{iE^{+}\left(  \tilde{y}_{1}+\tilde{y}%
_{2M+1}\right)  }\prod_{j=1}^{M}\left[  e^{iE^{+}\left(  \tilde{y}%
_{2j-1}+\tilde{y}_{2j+1}-2\tilde{y}_{2j}\right)  }\theta\left(  \tilde
{y}_{2j-1}-\tilde{y}_{2j}\right)  \theta\left(  \tilde{y}_{2j+1}-\tilde
{y}_{2j}\right)  \right]  \label{catalanintegral}%
\end{equation}
This is accompanied by a string of $2M$ consecutive $\delta$ factors
uninterrupted by $W$ factors, $\delta_{n_{1}n_{2}}\delta_{n_{2}n_{3}}%
\cdots\delta_{n_{2M}n_{2M+1}}$. We will prove in Section~\ref{subsec:catalan} that%

\begin{equation}
I_{M}\left(  E^{+}\right)  =\left(  \frac{i}{2E^{+}}\right)  ^{2M+1}C_{M}
\label{catalanint1}%
\end{equation}
where $C_{M}=\left(  2M\right)  !/\left[  M!\left(  M+1\right)  !\right]  $ is
the $M$th Catalan number.\cite{OEIS.A000108,*OEIS.A008315} The first few
Catalan numbers $0\leq M\leq5$ are $1$, $1$, $2$, $5$, $14$, $42$.

At this stage we can combine the $\left(  E^{+}\right)  ^{-1}$ factors in Eqs.~(\ref{ordrsint}) and (\ref{catalanint1}) with the $\tilde{E}_{2}$ or
$\tilde{\epsilon}$ factors. At each order there will be a single $\left(
E^{+}\right)  ^{-1}$ factor left unpaired, which gives the leading-log
renormalization $\delta D/D$ as the cutoff is reduced from $D$ to $D-\delta
D$. Collecting terms of all orders we see the S-matrix RG equation is of the
form of Eq.~(\ref{RPANazarovGlazman}), but the interaction $\Pi\left(
D\right)  $ is given by%

\begin{align}
\frac{\Pi_{jj^{\prime}}}{2}  &  =\frac{\alpha_{j}}{2}\delta_{jj^{\prime}%
}+\frac{\alpha_{j}}{2}\frac{\alpha_{j^{\prime}}}{2}\left\{  W_{jj^{\prime}%
}+\sum_{n_{1}}\frac{\alpha_{n_{1}}}{2}\left[  \delta_{jn_{1}}\delta
_{n_{1}j^{\prime}}+W_{jn_{1}}W_{n_{1}j^{\prime}}\right]  \right. \nonumber\\
&  +\sum_{n_{1}n_{2}}\frac{\alpha_{n_{1}}}{2}\frac{\alpha_{n_{2}}}{2}\left[
\delta_{jn_{1}}\delta_{n_{1}n_{2}}W_{n_{2}j^{\prime}}+W_{jn_{1}}\delta
_{n_{1}n_{2}}\delta_{n_{2}j^{\prime}}+W_{jn_{1}}W_{n_{1}n_{2}}W_{n_{2}%
j^{\prime}}\right] \nonumber\\
&  +\sum_{n_{1}n_{2}n_{3}}\frac{\alpha_{n_{1}}}{2}\frac{\alpha_{n_{2}}}%
{2}\frac{\alpha_{n_{3}}}{2}\left[  2\delta_{jn_{1}}\delta_{n_{1}n_{2}}%
\delta_{n_{2}n_{3}}\delta_{n_{3}j^{\prime}}+\delta_{jn_{1}}\delta_{n_{1}n_{2}%
}W_{n_{2}n_{3}}W_{n_{3}j^{\prime}}\right. \nonumber\\
&  \left.  +W_{jn_{1}}\delta_{n_{1}n_{2}}\delta_{n_{2}n_{3}}W_{n_{3}j^{\prime
}}+W_{jn_{1}}W_{n_{1}n_{2}}\delta_{n_{2}n_{3}}\delta_{n_{3}j^{\prime}%
}+W_{jn_{1}}W_{n_{1}n_{2}}W_{n_{2}n_{3}}W_{n_{3}j^{\prime}}\right]
+\cdots\Bigg\} \label{RPAvertex1}%
\end{align}
The rules to write down terms in Eq.~(\ref{RPAvertex1}) are as follows. At
$O\left(  \alpha^{m}\right)  $, there is a total number of $\left(
m-1\right)  $ factors of $\delta$ and $W$. The $\delta$ factors always appear
in even-length strings separated by the $W$ factors. Each string of $\delta$
of length $2M$ is associated with a multiplicative coefficient of the $M$th
Catalan number $C_{M}$. For instance, at $O\left(  \alpha^{17}\right)  $ there
is a term $W\delta\delta\delta\delta\delta\delta WWW\delta\delta\delta\delta
WW$, whose prefactor will be $C_{3}C_{2}=5\times2=10$.

We can resum Eq.~(\ref{RPAvertex1}) by observing that we can uniquely
construct every term containing a least one factor of $W$, by adding to an
existing term a (possibly empty) even-length string of $\delta$ followed by
one factor of $W$; e.g. the term $\delta\delta\delta\delta WW\delta\delta W$
is uniquely constructed as $\delta\delta\delta\delta$/$W$/$W\delta\delta W$.
In other words, $\Pi$ satisfies the relation%

\begin{equation}
\frac{\Pi_{jj^{\prime}}}{2}=\frac{\bar{\Pi}_{jj^{\prime}}}{2}+\sum_{l_{1}%
l_{2}}\frac{\bar{\Pi}_{jl_{1}}}{2}W_{l_{1}l_{2}}\frac{\Pi_{l_{2}j^{\prime}}%
}{2}\text{.} \label{RPAvertex2}%
\end{equation}
Here $\bar{\Pi}$ is the part of $\Pi$ which does not contain any factors of
$W$:%

\begin{align}
\frac{\bar{\Pi}_{jj^{\prime}}}{2}  &  =\frac{\alpha_{j}}{2}\delta_{jj^{\prime
}}+\frac{\alpha_{j}}{2}\frac{\alpha_{j^{\prime}}}{2}\left[  \sum_{n_{1}}%
\frac{\alpha_{n_{1}}}{2}\delta_{jn_{1}}\delta_{n_{1}j^{\prime}}+\sum
_{n_{1}n_{2}n_{3}}\frac{\alpha_{n_{1}}}{2}\frac{\alpha_{n_{2}}}{2}\frac
{\alpha_{n_{3}}}{2}2\delta_{jn_{1}}\delta_{n_{1}n_{2}}\delta_{n_{2}n_{3}%
}\delta_{n_{3}j^{\prime}}+\cdots\right] \nonumber\\
&  =\frac{\alpha_{j}}{2}\delta_{jj^{\prime}}\sum_{M=0}^{\infty}C_{M}\left(
\frac{\alpha_{j}}{2}\right)  ^{2M}=\frac{\alpha_{j}}{1+\sqrt{1-\alpha_{j}^{2}%
}}\delta_{jj^{\prime}}\text{.} \label{RPAvertex3}%
\end{align}
In the last line we have used the generating function of Catalan
numbers,\cite{OEIS.A000108}%

\begin{equation}
\sum_{M=0}^{\infty}C_{M}x^{M}=\frac{2}{1+\sqrt{1-4x}}\text{.}%
\end{equation}
Inserting Eq.~(\ref{RPAvertex3}) into Eq.~(\ref{RPAvertex2}) and solving for
$\Pi$, we obtain Eq.~(\ref{RPAvertex}) in the case of spatially uniform
interactions, $\alpha_{n}\left(  y\right)  =$ $\alpha_{n}$.

We now argue that the cutoff-dependence of the Luttinger parameter is through
Eq.~(\ref{alphaD}) as is the case with the first order calculation. To this
end, notice that it is values of $\tilde{y}_{n}\ $between $0$ and $O\left(
1/E^{+}\right)  $ that dominate the integral in Eq.~(\ref{catalanintegral}).
Therefore, when $D=\operatorname{Re}E^{+}\gtrsim v_{Fn}/L_{n}$, the integral
is governed by $v_{Fn}\tilde{y}_{n}\lesssim L_{n}$; in this range of
$\tilde{y}_{n}$, $\alpha_{n}\left(  v_{Fn}\tilde{y}_{n}\right)  =\alpha
_{n}\left(  0\right)  $. On the other hand, when $D\ll v_{Fn}/L_{n}$, the
integral is controlled mainly by $v_{Fn}\tilde{y}_{n}\gg L_{n}$, where
$\alpha_{n}\left(  v_{Fn}\tilde{y}_{n}\right)  =\alpha_{n}\left(
\infty\right)  $. This justifies the crossover behavior given by Eqs.~(\ref{alphaD}) and (\ref{scriptQ}), and concludes the calculation of the
self-energy terms in the RPA conductance.

Calculations of the RPA vertex corrections, or the ring diagrams, are
completely in parallel with the first-order vertex corrections except Eq.~(\ref{catalanintegral}) appears in the real space integrals. Here $E^{+}$ in
Eq.~(\ref{catalanintegral}) should be substituted for $\omega^{+}$. At the
$m$th order, all $m$ factors of $1/\omega^{+}$ in Eqs.~(\ref{ordrsint}) and
(\ref{catalanint1}) can be paired with the $m+1$ factors of $\omega^{+}$ from
loop energy integrals; the single unpaired $\omega^{+}$ will be combined with
the $1/\omega$ factor in Eq.~(\ref{Kubo}) so that the conductance is finite in
the DC limit. Also, all interaction strengths appearing here are those in the
leads $\alpha_{n}\left(  \infty\right)  $; this is because in the DC limit
$\omega\lesssim v_{Fn}/L_{n}$ for any lead $n$, and we may refer to our
argument in the previous paragraph for $D\ll v_{Fn}/L_{n}$. Eventually, taking
into account the dressing of the electron lines, we recover Eq.~(\ref{RPATLL}).

\subsection{Real space integral Eq.~\eqref{catalanint1}\label{subsec:catalan}}

To prove Eq.~(\ref{catalanint1}), we adopt the following change of variables
in Eq.~(\ref{catalanintegral}): $z_{0}=\tilde{y}_{1}$, $z_{2j-1}=\tilde
{y}_{2j-1}-\tilde{y}_{2j}$, $z_{2j}=\tilde{y}_{2j+1}-\tilde{y}_{2j}$, $1\leq
j\leq M$. The absolute value of the Jacobian of this change of variables is
simply $\left\vert \left(  -1\right)  ^{M}\right\vert =1$. We also introduce
the shorthand $s_{j}=\sum_{l=0}^{j}\left(  -1\right)  ^{l}z_{l}$. Eq.~(\ref{catalanintegral}) then becomes%

\begin{equation}
I_{M}\left(  E^{+}\right)  =\int_{0}^{\infty}dz_{0}\prod_{l=1}^{M}\left(
\int_{0}^{s_{2l-2}}dz_{2l-1}\int_{0}^{\infty}dz_{2l}\right)  \prod_{j=0}%
^{M}e^{2iE^{+}z_{2j}}%
\end{equation}
Now consider the auxiliary object,%

\begin{equation}
\tilde{I}_{M}\left(  E^{+},z_{0}\right)  \equiv\prod_{l=1}^{M}\left(  \int
_{0}^{s_{2l-2}}dz_{2l-1}\int_{0}^{\infty}dz_{2l}\right)  \prod_{j=0}%
^{M}e^{2iE^{+}z_{2j}}\equiv\left(  2iE^{+}\right)  ^{-2M}e^{2iE^{+}z_{0}}%
\sum_{l=0}^{M}\frac{T_{M,l}}{l!}\left(  -2iE^{+}z_{0}\right)  ^{l}
\label{Imtilde}%
\end{equation}
where $T_{M,l}$ are dimensionless coefficients; obviously $\tilde{I}%
_{0}\left(  E^{+},z_{0}\right)  =e^{2iE^{+}z_{0}}$ and $T_{0,0}=1$. $\tilde
{I}_{M}$ obeys the recurrence relation%

\begin{equation}
\tilde{I}_{M+1}\left(  E^{+},z_{0}\right)  =\int_{0}^{z_{0}}dz_{1}%
e^{2iE^{+}z_{1}}\int_{0}^{\infty}dz_{2}\tilde{I}_{M}\left(  E^{+},z_{0}%
-z_{1}+z_{2}\right)  \text{.} \label{Imtilderec}%
\end{equation}
Inserting Eq.~(\ref{Imtilde}) into Eq.~(\ref{Imtilderec}), we find that
$T_{M,l}$ satisfies the simple recurrence relation $T_{M+1,l}=\sum_{j=l-1}%
^{M}T_{M,j}$, and that $T_{M+1,0}=0$ ($M\geq0$). Such a recurrence relation
leads to the Catalan's triangle,\cite{OEIS.A008315}%

\begin{equation}
T_{M,l}=\frac{\left(  2M-l-1\right)  !l}{M!\left(  M-l\right)  !}\text{
}\left(  M\geq1\right)  \text{.}%
\end{equation}
Therefore,%

\begin{equation}
I_{M}\left(  E^{+}\right)  =\int_{0}^{\infty}dz_{0}\tilde{I}_{M}\left(
E^{+},z_{0}\right)  =-\left(  2iE^{+}\right)  ^{-2M-1}\sum_{l=0}^{M}T_{M,l}%
\end{equation}
Noting that $\sum_{l=0}^{M}T_{M,l}=C_{M}$, which is a property of Catalan's
triangle, we immediately recover Eq.~(\ref{catalanint1}).

\bibliography{YJunction}

\end{document}